# Optimal Sampling and Noise Mitigation for Vocal Fold Kinematics in High-Speed Videoendoscopy

Harikrishnan Unnikrishnan[1], Rita Patel[2,3], Kevin D. Donohue[4]

[1]Orchard Robotics, San Francisco, CA, USA

[2]Speech, Language, and Hearing Sciences, Indiana University, Bloomington, IN, USA

[3]Otolaryngology-Head and Neck Surgery, Indiana University School of Medicine, Indianapolis, IN, USA

[4]Electrical and Computer Engineering, University of Kentucky, Lexington, KY, USA

This work is accepted for Publication in Journal of Voice

This work was funded by National Institutes of Health (NIH) - National Institute on Deafness and Other Communication Disorders Grants R01DC017923 and R03DC11360. The authors gratefully acknowledge Dr. Dimitar Deliyski for providing 16000 fps High-Speed data.

## Abstract

**Objectives:** While the glottal area waveform is the most widely reported High-Speed Videoendoscopy derived signal for objective assessment of vocal fold kinematics, it conflates lateral fold excursion with longitudinal glottal opening, precluding direct estimation of velocity, acceleration, and left-right asymmetry. In contrast, displacement waveforms preserve these quantities but are more susceptible to spatial quantization and temporal sampling errors. We characterize these sampling errors and evaluate error reduction via Kalman smoothing and temporal interpolation. We further investigate the optimal frame rate bounds relative to pixel excursion, speed quotient and open quotient.

**Methods:** Sampling theory was used to derive analytical frame-rate bounds required to preserve kinematic feature fidelity based on temporal sampling and spatial resolution. Kalman smoothing was compared with Fast Fourier Transform-based denoising, combined with sinc or spline interpolation, to reduce sampling-related errors. Interpolation was evaluated using 4000 fps recordings (N=55) downsampled and reconstructed to the original rate. A 16000 fps dataset (N=12) served as the reference for noise-reduction evaluation. Waveforms were reduced to 2000–8000 fps, processed, and compared using peak displacement, velocity, open quotient, and speed quotient.

**Results:** Sinc interpolation yielded a lower median mean absolute error than spline (0.29 vs 0.52 pixels), with 85% of recordings (47/55) versus 45% (25/55) below 0.5 pixel. For kinematic feature estimation (N = 12), Kalman-sinc outperformed FFT-sinc in 75% of feature-by-frame-rate pairings.

**Conclusion:** Analysis reveals that temporal and spatial resolution must be considered together to avoid artifacts in computing kinematic features. Kalman-sinc preprocessing at 8000 fps is recommended for relative velocity and speed-quotient measures. While peak lateral displacement can be measured at 2000 fps, open quotient require minimal processing at 4000 fps for typical spatial resolution of 9 – 14 pixels of maximum excursion. We recommend reporting pixel excursion alongside kinematic measures to contextualize data reliability.

**Introduction**

While laryngeal videostroboscopy (LVS)[1,2,3] remains the standard in clinical practice, high-speed videoendoscopy (HSV)[4,5] provides the temporal resolution necessary to directly quantify vocal fold kinematics — including velocity, acceleration, and left-right asymmetry — that stroboscopic reconstruction cannot resolve without additional assumptions.[6,7] HSV therefore offers a unique opportunity for objective, model-independent assessment of vocal fold biomechanics in both research and clinical settings.

Although the glottal area waveform (GAW) is widely reported in HSV studies,[8,9,10,11,12] it conflates vocal fold lateral displacement with glottal opening area. Consequently, it does not directly support relative vocal fold velocity, relative acceleration, or left–right asymmetry. Displacement waveforms[13,14], kymograms extracted from Digital Kymography (DKG)[15,16,17] and hardware-driven Videokymography (VKG)[18,19,20,21] are more suited for these analyses. While displacement waveform and digital kymograms are derived from HSV recordings, VKG natively captures one horizontal line per sampling instant (typically 7,200–8,000 Hz) to yield similar waveform, but with reduced digital storage and lighting requirements. While kymograms (DKG or VKG) display the temporal evolution of pixel intensity along a fixed scan line across the glottis[22,23], displacement waveforms track the left and right vocal fold edge across successive frames, typically at a location midway between the anterior and posterior commissures.[13,14] Both representations encode the temporal dynamics of vocal fold motion and can be used to estimate kinematic quantities such as velocity, acceleration, and left-right asymmetry. However, unlike GAW, both are more directly affected by spatial quantization and temporal undersampling because their measurements depend on accurate localization of tissue boundaries over time.

This paper focuses on displacement-based kinematics as the displacement waveform provides a direct estimate of vocal fold edge motion after compensating for vocal fold rotation and endoscopic movement, resulting in a more geometrically accurate representation of tissue displacement. The displacement waveforms are used to derive clinically relevant kinematic measures such as open quotient, speed quotient, peak lateral displacement, and relative opening/closing velocities.[13,14] The Figure 1 illustrates the displacement waveform and its associated cycle timing parameters, including period, open phase, opening phase, closing phase, and peak amplitude excursion.

The selection of an optimal HSV frame rate ($f_s$) has historically been guided by empirical observation rather than theoretical derivation. Deliyski et al.[4] demonstrated that while 2000 fps may be sufficient for fundamental frequency estimation, it is inadequate for capturing the fine temporal details of the mucosal wave, particularly in pathological or high-pitched voices. Consequently, clinical and research systems have moved toward adopting higher rates, with 4000 fps becoming a commonly used benchmark.[8,10,24]

Several studies have demonstrated that quantitative HSV measurements depend on acquisition parameters. Schützenberger et al.[25] reported that approximately 90% of computed HSV parameters change with frame rate. However, the underlying mechanism driving this dependence remains unclear. In parallel, the impact of the spatial resolution, specifically, the number of pixels representing the vocal fold displacement ($K$) is often overlooked. Schlegel et al.[26] showed that spatial camera resolution substantially influences HSV-derived parameters: reducing resolution from 256×256 to 128×64 pixels caused statistically significant changes in 51–63% of glottal area waveform parameters across sexes. Because lower resolution increases the pixel size, and thus the quantization step, these findings are consistent with spatial quantization contributing to parameter uncertainty.

The effect of temporal data length has also been investigated. Schlegel et al.[27] reported that only 3 of 16 perturbation parameters (18.8%) from the Glottis Analysis Tools (version 2018) were significantly influenced by the number of the analyzed cycles and recommend a minimum of 20 consecutive vibratory cycles for stable parameter estimation. Despite these advances, the field lacks a unified framework linking temporal sampling requirements with spatial resolution. This study addresses this gap by deriving analytical frame rate bounds for displacement-based kinematic measurements and demonstrating that the benefit of increased temporal resolution is fundamentally constrained by available spatial resolution.

The displacement waveforms are affected by two primary digital sampling effects. First, *spatial quantization* occurs because the optical image of the vocal folds is projected onto a discrete pixel array, where continuous tissue motion is represented by finite spatial sampling units. In digital signal processing (DSP), this is a well-known quantization effect that introduces an inherent noise floor due to finite spatial resolution[28]. Second, *temporal sampling* is limited by the camera frame rate. If the number of frames per vibratory cycle is too low, peak lateral displacement and opening/closing relative velocity *estimates* are systematically underestimated (or aliased) and timing measures (e.g. open quotient, speed quotient) become unreliable.[25]

Although increasing frame rate and spatial resolution can reduce these errors, practical constraints of sensor cost, light exposure, storage, etc. prevent unlimited increases in sampling. Crucially, when temporal frame rates are elevated without a proportional increase in spatial resolution, the minute inter-frame tissue movement drops below the sensor grid resolution. This creates artificial step-discontinuities that dominate the displacement waveform. The staircase effect observed in displacement waveforms is closely related to the quantization and staircase problems during DSP data capture.

Recovering the physical trajectory from such quantized measurements is a recognized signal reconstruction problem, for which temporal estimation and smoothing methods can be used to infer sub-resolution motion[28]. Therefore, an optimal acquisition strategy requires balancing temporal and spatial resolution. This paper addresses this challenge by characterizing the relationship between frame rate, region-of-interest spatial resolution, and vocal fold cycle frequency. We derive the minimum (min) and maximum (max) frame rates as a function of frequency, cycle asymmetry (open quotient and speed quotient), samples per vibratory phase and pixel excursion. We then evaluate processing pipelines, constructed from standard signal processing methodologies conventionally deployed to handle quantization noise and grid artifacts[28]. Specifically, we compare two foundational noise reduction approaches—Kalman smoothing[29], novel in this context for quasi-periodic vocal fold displacement signals, and Fast Fourier Transform (FFT)-based denoising[Error! Reference source not found.]—each combined with spline or sinc interpolation, to identify the optimal processing strategy, including no-processing, for improving kinematic feature estimation.

The contributions of this paper are: (i) analytical/theoretical derivation of the minimum (min) and maximum (max) frame rate bounds $f_{s,min}$ and $f_{s,max}$ respectively, as a function of fundamental frequency, open quotient, speed quotient, pixel excursion and required samples per vibratory phase to limit vibration parameter estimation errors; and (ii) the application of a signal processing framework for kinematic feature estimation. We evaluate Kalman smoothing on quasi-periodic vocal fold displacement waveforms and compare it with FFT-based denoising, combined with temporal interpolation methods, to identify processing strategies that improve the fidelity and reliability of extracted kinematic features.

## 2. Materials and methods

### 2.1. Theoretical analysis

In this section, we first characterize the theoretical noise floor in terms of Signal-to-quantization-Noise Ratio (SNR) of the displacement waveform. SNR highlights the expected fidelity of the displacement waveform relative to the actual physical motion. We next define the lower (Minimum (min), $\boldsymbol{f_{s,min}}$) and upper bounds (Maximum (max), $\boldsymbol{f_{s,max}}$) of the sampling frame rate ( $\boldsymbol{f_s}$) and demonstrate the strong interdependence between spatial resolution and temporal frame rate requirements. Here $\boldsymbol{f_{s,min}}$ denotes the minimum rate required to reliably estimate derivative features (velocity, acceleration and speed quotient), whereas $\boldsymbol{f_{s,max}}$ represents the frame rate corresponding to a specified pixel excursion ($K$), beyond which additional sampling results in oversampling. Finally, we present signal processing techniques that mitigate spatial and temporal sampling noise, thereby improving the accuracy of kinematic feature estimation.

#### *2.1.1. Derivation of Signal- to-Quantization-Noise Ratio (SNR)*

Consider the vocal fold edge lateral displacement, measured relative to the line connecting the anterior and posterior commissures and expressed in pixel units. Let ($K$) denote the peak glottal excursion expressed in pixels. Equivalently, ($K$) represents the number of quantization levels spanning the peak vibratory displacement. With a quantization step size $\Delta = 1$ pixel, and the adaptive thresholding localizing vocal fold edge to the nearest whole pixel, the quantization error is uniformly distributed over the interval $[-0.5, 0.5]$. The analytical variance of this continuous uniform random distribution yields a noise power (variance) of $\sigma_e^2 = \Delta^2/12 = 1/12$.[31] Approximating the vocal fold displacement as a sinusoidal signal with zero-to-peak amplitude ($K$), the corresponding signal power is $\sigma_x^2 = K^2/2$.[31] Therefore, the theoretical signal-to-quantization-noise ratio (SNR) is:

$$SNR = 10\log_{10}\left(\frac{K^2/2}{1/12}\right) = 20log_{10}(\sqrt{6}\,K).$$

In our clinical dataset, the peak glottal excursion ($K$) typically ranges from 7-10 pixels in children, 8-12 pixels in females, and 9 – 14 pixels in males. These values correspond to theoretical SNRs of approximately $26 - 31\ dB$, indicating that the displacement waveform is inherently limited by spatial quantization noise. Because differentiation amplifies high-frequency noise, this spatial noise floor becomes substantially more pronounced in derivative signals, reducing the accuracy of kinematic features such as velocity and acceleration. Consequently, signal processing is required to suppress quantization noise before reliable feature extraction can be performed.

#### *2.1.2. Derivation of Frame Rate bounds*

To accurately estimate the first and second-order derivatives of the displacement waveform (i.e. velocity and acceleration), sufficient number of samples within each vibratory phase is required. Let ($N$) denote the minimum number of frames per phase required for reliable derivative estimation. The minimum frame rate ($f_{s,min}$), is therefore determined by ensuring that even the shortest vibratory phase is sampled by at least ($N$) frames.

Derivative estimation is performed using the 3-point central difference[32] method because it provides a more accurate and less noise-sensitive estimate than forward or backward difference alone. The 3-point central difference method requires one sample on each side of the evaluation point. To obtain at least one valid interior derivative estimate while maintaining a one-sample margin at each phase boundary, each vibratory phase must contain a minimum of five samples ($N = 5$). Because the actual vocal fold motion is inherently nonlinear, this represents a theoretical lower bound; additional samples per phase further improve the accuracy of derivative estimation.

This theoretical requirement is consistent with previous empirical observations. Deliyski et al.[24] empirically established minimum frame rates of 4000 fps for women and 3200 fps for men to enable reliable subjective evaluation of vocal fold vibration. Using representative fundamental frequencies of $f_0 = 220$Hz for women and $f_0 = 120$ Hz for men, together with typical values of Open Quotient ($O_q = 0.6$) and Speed Quotient ($S_q = 1.0$), these recommended frame rates correspond to at least ($N \geq 5$) five frames within the shortest vibratory phase. Thus, the empirical analysis agrees with the theoretical analysis of the minimum sampling requirement derived from the 3-point central difference method. Therefore, the frame rate must be sufficiently high to ensure that even the shortest phase of the glottal cycle is sampled by at least ($N$) frames.

The opposite limitation occurs when the frame rate is excessively high. As the sampling interval decreases, the true lateral displacement of the vocal folds between consecutive frames may become smaller than the image resolution (1 pixel). Under this condition, the measured displacement waveform becomes increasingly dominated by staircase quantization artifacts rather than the underlying continuous motion. An upper bound on the frame rate ( $f_{s,max}$), can therefore be established by requiring that the average lateral displacement between consecutive frames exceeds one pixel. The average velocity, rather than the maximum velocity, is used to derive this criterion. Enforcing a one-pixel displacement criterion at the instant of maximum velocity would satisfy the requirement at a single point in the vibratory cycle, while allowing sub-pixel motion during the slower portions of the opening and closing phases, where quantization error would become dominant. Using the average velocity ensures that quantization effects remain limited throughout the entire phase.

The opening ($T_{opening}$) and closing ($T_{closing}$) durations ($T$) (seconds) of the glottal cycle can be expressed in terms of the open quotient ($O_q$), speed quotient ($S_q$), and fundamental frequency ($f_0$) as below (See Appendix A for the detailed derivation):

$$T_{opening} = \frac{O_q S_q}{(S_q + 1) \times f_0}$$

$$T_{closing} = \frac{O_q}{(S_q + 1) \times f_0}$$

where $f_0$ is the fundamental frequency, $O_q$ is the open quotient, $S_q$ is the speed quotient, $K$ is the peak pixel excursion. Applying the minimum sampling requirement, the frame rate must be sufficiently high to provide at least ($N$) samples within the shortest vibratory phase: $f_s \geq \frac{N}{T_{short}}$ where $T_{short}$ is the duration of the shorter opening and closing phases. Applying the upper sampling requirement, the average lateral displacement between two consecutive frames must exceed one pixel. These two constraints yield the following analytical expressions for the minimum and maximum frame rates:

$$f_{s,min} = \frac{N(S_q + 1)\, f_0}{O_q \min(S_q,\ 1)}$$

and

$$f_{s,max} = \frac{K f_o (1 + S_q)}{O_q \max(S_q, 1)}$$

where ($N$) is the minimum number of frames in the shortest vibratory phase, ($K$) is the peak lateral excursion (pixel), ($O_q$) is the open quotient, ($S_q$) is the speed quotient, and ($f_0$) is the fundamental frequency. The complete derivation is provided in Appendix A.

The $\min(S_q,\ 1)$ term automatically selects the shorter glottal phase. When $S_q < 1$, the opening phase is shorter and contributes the factor $S_q$, and when $S_q > 1$, the closing phase is shorter and contributes a factor of 1. Thus, the formula for ($f_{s,min}$) is valid over the full physiological range of $S_q$. Similarly, the $\max(S_q,\ 1)$ selects the longer phase in the formula for ($f_{s,max}$). Together, ($f_{s,min}$) and ($f_{s,max}$) define the admissible operating range for the recording frame rate:

$$f_{s,min} \leq f_s \leq f_{s,max}$$

For such an operating range to exist, the lower bound must not exceed the upper bound. Substituting the expressions for $f_{s,min}$ and $f_{s,max}$ in this inequality yield the following operating condition (See Appendix A for the step-by-step derivation):

$$N \leq K \min(S_q, S_q^{-1})$$

This inequality defines the feasibility condition for accurate kinematic measurements. Specifically, the minimum sampling requirement for reliable derivative estimation ($N \geq 5$) for accurate derivatives can only be satisfied if the available spatial resolution, represented by peak pixel excursion $(K)$, is sufficiently large. The parameter $(N)$ is directly regulated by selecting recording frame rate $(f_s)$, whereas $(K)$ scales with the spatial resolution of the vocal fold motion within the image. Increasing pixel excursion $(K)$, can be achieved by increasing the image magnification—for example, by positioning the endoscope closer to the vocal folds or ensuring that the region of interest occupies a greater proportion of the image. Because the working distance and optical zoom are typically constrained in clinical practice, $(N)$ and $(K)$ are most practically optimized by adjusting the frame rate and the image spatial resolution, respectively.

The inequality also demonstrates a fundamental coupling between temporal and spatial sampling. Increasing the frame rate raises the number of samples per vibratory phase but simultaneously decreases the inter-frame displacement. Consequently, the frame rate cannot be increased beyond $(f_{s,max})$ unless it is accompanied by a proportional increase in pixel excursion (or equivalently, spatial resolution). Otherwise, the measured displacement waveform becomes dominated by staircase quantization artifacts rather than true vocal fold motion. Figure 2 illustrates this effect. At a sampling rate of 16000 fps, the displacement waveform exhibits a pronounced staircase pattern, with consecutive samples remaining at the same pixel value before abruptly changing to the next level. This quantization artifact obscures the underlying continuous motion and degrades the accuracy of velocity and acceleration estimates. The problem is particularly important in pediatric and other high-pitched voices, where the higher fundamental frequency increases $(f_{s,min})$. If the available pixel excursion is insufficient, the lower and upper frame-rate bounds may no longer overlap, leaving no feasible operating region regardless of the selected frame rate.

In summary, $(f_{s,min})$ defines the lowest acceptable frame rate required to obtain sufficient samples per vibratory phase for accurate derivative estimation, whereas $(f_{s,max})$ defines the highest acceptable frame rate before quantization artifacts dominate the displacement waveform. The operating condition above ensures that these two bounds are compatible for the available spatial resolution. Having established these analytical limits and the intrinsic sampling noise floor, the following section evaluates signal-processing techniques designed to mitigate spatial and temporal sampling noise and improve the fidelity of extracted kinematic features. The overall processing pipeline is presented first, followed by a detailed description of each component.

### 2.2 Signal Processing pipeline

The proposed signal processing pipeline is illustrated in Figure 3. First, raw lateral displacement waveforms are extracted from HSV frames by applying adaptive thresholding within a motion-based region of interest.[33,34] which localizes the vocal fold edge to the nearest whole pixel, The extracted displacement waveform are then denoised using either Kalman smoothing[29] or FFT-based denoising[30] to reduce quantization and discretization noise, while preserving the underlying vocal fold motion. The denoised waveform is subsequently interpolated using either sinc or spline interpolation to improve the temporal resolution and enable sub-frame localization of kinematic events. Finally, the glottal opening and closing instants are estimated with sub-sample precision by using linear extrapolation of the smoothed/interpolated displacement waveform (Appendix B). The experiments presented in the following sections evaluate the accuracy of alternative denoising and interpolation methods and evaluate their impact on the accuracy of the extracted kinematic measures of interest.

### 2.2.1 Methods for reducing Quantization noise

*Kalman Smoothing*

To reduce quantization noise and address the unique kinematic characteristics of HSV, we employ a Kalman smoother.[29] The smoother estimates the underlying vocal fold motion by optimally combining a physics-based motion model with the observed displacement measurements, thereby suppressing measurement noise while preserving the true dynamics (Appendix B).

The proposed implementation incorporates three adaptations specifically designed for HSV data: First, the state vector includes jerk (the third derivative of displacement) to capture the rapid changes in acceleration associated with vocal fold impact during collision. Second, a dedicated closed-phase model suppresses spurious zero-valued observations that occur during complete glottal closure. Third, the observation noise covariance is adaptively adjusted according to the expected inter-frame displacement, allowing the filter to account for the dependence of quantization error on frame rate.

In this framework, the method assumes that the true vocal fold lateral motion is temporally smooth and that the observed discontinuities arise primarily from spatial quantization and edge-detection errors. This modeling is augmented by bidirectional smoothing which further improves state estimation by incorporating information from both past and future observations, while the specialized treatment of glottal closure preserves kinematic continuity across closed phases. The inclusion of jerk is critical because it allows the model to capture the rapid force changes during vocal fold collision while successfully distinguishing genuine motion from stochastic jumps caused by pixel-level tracking errors.

When the vocal folds are completely closed, no glottal edge is detected, and displacement is set to zero by the tracking algorithm. These zero-valued measurements do not represent the true vocal fold position and are therefore treated as missing observations rather than valid measurements. During the closed phase, the Kalman smoother relies solely on the motion model to propagate the state estimate, allowing the predicted trajectory to evolve continuously without being biased by the spurious zero observations (Appendix B). Once the folds reopen and valid displacement measurements become available, the filter resumes normal measurement updates from the predicted state. This approach preserves kinematic continuity across glottal closure and prevents the ringing[35] and overshoot artifacts commonly produced by spline interpolation or Butterworth filtering when they encounter the abrupt amplitude discontinuity associated with complete glottal closure.

Once the left and right displacement waveforms are estimated, the opening and closing instants are detected with sub-frame precision by linearly extrapolating their difference signal. For each transition, the sample spanning the zero-crossing is discarded to avoid bias from the discontinuity. Instead, the two adjacent samples within the open phase— immediately preceding the closing instant or immediately following the opening instant—are used to linearly extrapolate the difference signal to its zero crossing. This yields sub-frame estimates of the opening and closing instants required for accurate computation of the open quotient and speed quotient.

The state-space equations underlying the Kalman smoothing, together with the HSV- specific adaptations described above, are provided in Appendix B. In summary, the Kalman smoother reconstructs a physically plausible, continuous displacement trajectory from the quantized, staircase-like measurements, enabling reliable estimation of velocity and acceleration.

*FFT-based denoising*

Fast Fourier Transform (FFT) denoising[30] provides a non-parametric alternative to Kalman filtering by separating the vocal fold displacement signal from the quantization noise in the frequency domain. The displacement waveform is quasi-periodic, with most of its energy concentrated at the fundamental frequency $f_0$ and its harmonics. The energy of this signal is distributed over limited spectral representation. In contrast, the quantization noise arises from pixel round-off error and is well modeled as spectrally white, distributing its energy broadly across the frequency spectrum. This difference in spectral characteristics enables effective denoising by thresholding on the frequency-domain coefficients, preferentially suppressing broadband noise while preserving the dominant spectral components associated with vocal fold motion.

Let $R[n]$ denote the Discrete Fourier Transform (DFT) coefficients of the raw displacement waveform computed over several contiguous cycles. Hard thresholding is applied to get the denoised spectrum ($\hat{R}[k]$):

$$\hat{R}[k] = R[k] if |R[k]| \geq T_Q, 0 \text{ otherwise}$$

and the denoised waveform is recovered as $\hat{x}[n] = IDFT\{\hat{R}[n]\}$ where $IDFT$ denotes the Inverse Discrete Fourier Transform operation. The threshold $\left(T_Q\right)$ is set adaptively as three times the median magnitude of the upper half of the spectrum where little to no signal harmonics are present and the spectrum is dominated by noise.Error! Reference source not found. This data-driven rule avoids the need for a separate noise estimate and is robust to variations in $f_0$ and cycle count.

## 2.2.2. Interpolation for temporal refinement

Following denoising, the displacement waveform is interpolated to increase its temporal resolution and further reduce sampling-related errors in the estimation of kinematic features. Interpolation is performed after the denoising stage (Kalman smoothing or FFT-based denoising, depending on the pipeline variant under evaluation), and the two interpolation methods are evaluated: cubic spline interpolation and sinc interpolation. Cubic spline interpolation fits a third-degree polynomial across neighboring samples to ensure piecewise continuity, providing a smooth curve for the interpolated waveform. By leveraging local signal information, this technique is essential for robustly calculating kinematic derivatives—such as velocity and acceleration—without introducing artificial, abrupt changes that could be misinterpreted as pathological motion. In contrast to cubic spline interpolation, sinc interpolation utilizes the global information from the entire quasi-periodic waveform. By convolving the signal with a band-limited sinc function, this technique is theoretically optimal for reconstructing periodic waveforms. For the approximately periodic vocal fold displacement waveform, this approach preserves the spectral content of the glottal cycle and minimizes interpolation error, making it theoretically optimal when the sampling assumptions are satisfied.

## 2.3 Experimental design

### 2.3.1 Datasets

Two independent datasets were used to evaluate the proposed framework, each serving a distinct purpose.

*Dataset 1 (4000 fps, interpolation evaluation)*. The first dataset consists of 4000 fps, $512 \times 256$ pixels HSV recordings of habitual sustained phonation from 55 vocally healthy subjects (27 children, 17 females, and 11 males) collected at the Vocal Physiology and Imaging Laboratory, University of Kentucky[13]. The cohort spans children and adults of both sexes, providing a broad range of vocal-fold morphologies, phonation characteristics, fundamental frequencies, and pixel excursions. This diversity makes the dataset well suited for evaluating the performance of interpolation methods under a wide range of physiological conditions. All subjects were free of vocal fold lesions and dysphonia and were enrolled after completing the Institutional board consent and assent forms. Adults with a history of smoking were excluded from the study.

*Dataset 2 (16000 fps, kinematic-feature evaluation)*. The second dataset comprises of habitual sustained phonation from 12 vocally healthy subjects (6 males and 6 females) recorded at the University of South Carolina[24]. Recordings were obtained using a high camera sensitivity (ISO 4000) and a spatial resolution of $639 \times 639$ pixels, thereby minimizing the contribution of image sensor noise to the measured displacement signals. This dataset was selected as the reference standard for evaluating kinematic feature accuracy because its substantially higher temporal resolution provides a closer approximation to the underlying continuous vocal fold motion. As reported by Deliyski et al.[24], frame rates above 4000 fps are necessary for reliable vocal fold kinematic assessment, while the temporal sampling error associated with 4000 fps recordings has not been fully characterized. Consequently, the 16000 fps recordings provide an appropriate reference against which the performance of lower-frame-rate recordings and the proposed signal-processing methods can be evaluated.

### 2.3.2 Interpolation Methods Evaluation

Interpolation performance was evaluated using Dataset 1 (55 subjects, 4000 fps). Each displacement waveform was first pre-smoothed with FFT-based denoising to suppress pixel quantization artifacts and produce a stable reference signal. This preprocessing denoising step was performed solely to isolate interpolation performance by minimizing the influence of measurement noise. The comparative performance of FFT denoising versus Kalman smoothing are evaluated separately in Section 2.3.3. The pre-smoothed waveform was downsampled from 4000 fps to 2000 fps by retaining every other sample. Sinc and cubic spline interpolation were subsequently applied independently to reconstruct the waveform on the original 4000 fps sampling grid.

Interpolation accuracy was quantified using the mean absolute error (MAE, in pixels) computed only at the interpolated samples (i.e., the odd-indexed samples omitted during downsampling). The median error was evaluated exclusively during the glottal opening phase, where the displacement changes most rapidly and interpolation errors are expected to be greatest. For each recording, the median MAE was calculated separately for the left and right vocal fold displacement waveforms to reduce sensitivity to localized tracking errors, and the average of the two median values was reported as a single summary value per recording. Results were then aggregated across 55 subjects (17 females, 11 males, 27 children).

The downsampling procedure used in this evaluation retained every other sample and therefore does not replicate the finite integration time of a real lower-frame-rate camera. Consequently, the absolute error magnitudes may therefore differ from those obtained using true 2000 fps recordings. However, because both interpolation methods were evaluated using the identical downsampled signals, the relative comparison between the interpolation methods remains unbiased.

### 2.3.3 Noise Reduction Methods Evaluation

The effectiveness of the proposed noise reduction methods (Kalman smoothing and FFT denoising) was evaluated using Dataset 2 (12 subjects, 16000 fps) by quantifying their impact on the accuracy of extracted kinematic features at different recording frame rates.

To simulate lower-frame-rate acquisitions, the original 16000 fps recordings were temporally averaged by combining consecutive groups of ($M$) frames ($M = 2, 3, 4$, and 8), corresponding to effective frame rates of 8000, 5333, 4000, and 2000 fps respectively. Unlike the simple sample decimation using in the interpolation evaluation in Section 2.3.2 (Dataset 1), frame averaging more closely approximates the finite integration time of a camera operating natively at a lower frame rate. The vocal fold lateral displacement waveform was subsequently extracted from each frame-averaged image sequence.

Seven processing pipelines were evaluated at each frame rate: (i) no processing, (ii) sinc interpolation only, (iii) spline interpolation only, (iv) Kalman smoothing followed by sinc interpolation, (v) Kalman smoothing followed by spline interpolation, (vi) FFT denoising followed by sinc interpolation, and (vii) FFT denoising followed by spline interpolation.

For the no-processing condition, kinematic features were computed directly from the frame-averaged waveform at its native sampling rate. For all other pipelines, the displacement waveform was first upsampled to the original 16000 fps temporal grid using either sinc or spline interpolation before feature extraction.

The extracted kinematic features[13] included Open quotient (OQ), Speed Quotient (SQ), Peak Lateral Displacement (PLD), Average Relative Opening Velocity (AOV), Peak Relative Opening Velocity (POV), Average Relative Closing Velocity (ACV) and Peak Relative Closing Velocity (PCV). Relative velocity was computed as the finite-difference derivative of the lateral displacement waveform. These features span the three categories of displacement waveform-based kinematics: timing (OQ, SQ), amplitude (PLD), and velocity (AOV, POV, ACV, PCV).

For each processing pipeline, the extracted features were compared with those obtained from the corresponding 16,000fps recording processed independently using Kalman smoothing, which served as a reference. Performance was quantified using the Mean Absolute Percentage Error (MAPE).[34] Left and right vocal fold measurements were analyzed separately and then pooled to summarize overall error trends.

Perturbation measures such as jitter and shimmer were not evaluated because they quantify cycle-to-cycle variability rather than kinematic properties derived directly from the displacement waveform. Furthermore, because this study is intended to validate the analytically derived sampling framework rather than establish normative population values, and because the 16000 fps dataset is necessarily limited in size, results are reported using descriptive error statistics throughout.

### 2.3.4 Reference method

A key premise of this study is that raw HSV recordings do not constitute ground truth. Even at 16000 fps, the measured vocal fold displacement remains limited by the finite spatial resolution of the imaging system, and the extracted displacement is therefore affected by pixel quantization noise. Because no independently validated ground-truth displacement data exists for these recordings, an alternative reference standard is required.

Accordingly, the Kalman-smoothed displacement waveform obtained from the 16000 fps recordings was adopted as a reference signal. Among all frame rates evaluated, the 16000 fps recordings provide the highest temporal sampling density and, after Kalman smoothing, offer a physically plausible estimate of the underlying continuous vocal fold motion. Using a common 16,000 fps reference enables all lower-frame-rate pipelines to be assessed against a plausible reference displacement trajectory. To ensure a fair and controlled comparison between denoising strategies, both processing strategies (Kalman and FFT) were evaluated on identical downsampled input trajectories and both utilized identical sinc/spline interpolation for reconstruction. This design isolates the denoising mechanism as the sole experimental variable.

It should be noted that this reference is necessarily self-referential for the 16,000 fps Kalman-smoothed pipelines themselves. Consequently, the Kalman–Sinc and Kalman–Spline pipelines at 16,000 fps exhibit near zero error by construction. For this reason, these results are not used to formulate any recommendations. Instead, all quantitative recommendations presented in Table 4 are based exclusively on the independent evaluations performed at 2000–8000 fps, where the reference remains distinct from the processing pipeline under investigation.

For each kinematic feature, performance was quantified using the Mean Absolute Percentage Error (MAPE) as:

$$MAPE = \frac{1}{M}\sum_{m=1}^{M}\left|\frac{\hat{f}_m - f_m}{f_m}\right| \times 100$$

where $\hat{f}_m$ is the feature value from the pipeline under evaluation, $f_m$ is the corresponding Kalman-smoothed 16000 fps reference value, and $M$ is the number of subjects. Thus, the reported MAPE quantifies how accurately each processing pipeline recovers displacement-derived kinematic features from lower-frame-rate, spatially quantized recordings relative to the best available estimate of the underlying vocal fold motion.

## 3. Results

### 3.1 Interpolation quality: sinc vs. spline (N = 55)

Figure 4 presents box-and-whisker plots comparing the mean absolute error (MAE) of sinc and spline interpolation across 55 subjects in Dataset 1. Sinc interpolation demonstrated consistently lower reconstruction errors than spline interpolation, with a median MAE of 0.29 pixels compared with 0.52 pixels for spline interpolation. The upper quartile (Q3) values further demonstrate this difference, with 75% of sinc interpolation errors remaining below 0.39 pixels, whereas 75% of spline interpolation errors remained below 0.76 pixels.

This improvement was also reflected in the proportion of subjects achieving sub-pixel reconstruction accuracy. Sinc interpolation produced an MAE below 0.5 pixels in 85% of subjects (47/55), whereas spline interpolation achieved the same criterion in only 45% of subjects (25/55). The 75th percentile of the sinc interpolation error distribution (0.39 pixels) was lower than the median error of spline (0.52 pixels), indicating that sinc interpolation provided systematically more accurate waveform reconstruction across the cohort.

### 3.2 Full pipeline analysis (N = 12)

The full pipeline evaluation demonstrated clear differences in sensitivity among kinematic feature types. Peak Lateral Displacement (PLD) was robust to frame-rate reduction and required no additional processing across all evaluated frame rates. In contrast, timing- and velocity-based features exhibited greater dependence on the processing pipeline. Open Quotient (OQ) was effectively stabilized through interpolation alone, whereas Speed Quotient (SQ) and velocity-derived features required the complete processing pipeline, consisting of Kalman smoothing followed by sinc interpolation, to achieve reliable estimation.

#### 3.2.1 Peak Lateral Displacement Features

PLD features had consistently low errors with Mean Absolute Percentage Error (MAPE) values remaining within 7–10% (Table 1) across all the evaluated frame rates. The no-processing condition produced consistently low errors. This outcome is expected, as zero-derivative displacement metrics such as PLD do not depend on fine frame-to-frame temporal derivative estimation or velocity continuity.

#### 3.2.2 Quotient Features

Quotient-based features demonstrated increased sensitivity to temporal sampling and signal-processing strategy, particularly at lower frame rates. At 2000 fps, the Open Quotient (OQ) exhibited a MAPE of 12.96% in the no-processing condition. Temporal interpolation alone effectively mitigated this sampling limitation, with Sinc-only and Spline-only interpolation reducing

OQ MAPE to 4.93% and 5.37%, respectively (Table 1). Applying Kalman smoothing prior to sinc interpolation produced a comparable MAPE of 6.20%. Because OQ estimation depends primarily on determining temporal threshold crossings during glottal opening and closing, high-order temporal reconstruction alone recovers sub-sample event timing without filter-induced smoothing of boundary transitions.

In contrast, the Speed Quotient (SQ) showed substantially greater sensitivity to sampling and processing effects. At 2000 fps, the no-processing condition resulted in a MAPE of 70.39%, whereas the complete Kalman–sinc pipeline reduced the error to 16.57%. These results indicate that SQ estimation is considerably more dependent on noise suppression and temporal reconstruction than OQ estimation, likely due to its greater reliance on accurate characterization of the relative timing of opening and closing dynamics.

### 3.2.3 Velocity Features

Velocity-derived features exhibited the greatest sensitivity to both temporal sampling and noise-processing strategy. When combined with Kalman smoothing, both cubic spline and sinc interpolation effectively stabilized average relative velocity estimates. At 8000 fps, the Kalman–spline pipeline achieved MAPEs of 8.29% for Average Relative Opening Velocity (AOV) and 15.57% for Average Relative Closing Velocity (ACV), while the Kalman–sinc pipeline achieved comparable MAPEs of 8.91% and 15.62%, respectively (Table 2). Estimation errors reached a stable level at 8000 fps when processed with the full Kalman–sinc pipeline, suggesting that further increases in frame rate provide limited additional benefit.

Peak velocity features were substantially less stable without noise reduction. In the no-processing condition, Peak Relative Opening Velocity (POV) and Peak Relative Closing Velocity (PCV) exhibited increasing errors, with PCV reaching a maximum MAPE of 242.8% at 16,000 fps (Table 3, Figure 6). These results indicate that increasing temporal resolution alone does not eliminate instability in derivative-based features, as high-frequency quantization noise is amplified during velocity estimation. At 8000 fps, sinc interpolation without denoising resulted in MAPEs of 45.3% for POV and 78.4% for PCV. Furthermore, sinc-only interpolation at 16,000 fps still produced a PCV MAPE of 231%. In contrast, the complete Kalman–sinc pipeline substantially reduced and stabilized peak velocity errors, achieving MAPEs of 15.4% for POV and 22.0% for PCV at 8000 fps.

Together, these findings demonstrate that derivative-based kinematic features are limited not only by temporal sampling density but also by spatial quantization noise. Noise suppression prior to differentiation is therefore essential for reliable estimation of peak velocity metrics.

### 3.2.3 Direct Comparison: Kalman vs. FFT Denoising

A direct comparison of the two noise reduction strategies (Figure 7) showed that the Kalman-sinc pipeline generally provides more accurate kinematic feature estimation than the FFT-sinc pipeline. Figure 7 illustrates this relationship by plotting the MAPE of Kalman-Sinc on the vertical axis and that of the FFT-Sinc on the horizontal axis, where greater distance from the dashed identity line indicates a larger performance divergence between the two approaches. Because sinc interpolation was previously established as the optimal temporal interpolation method (Section 3.1), only the Kalman–sinc and FFT–sinc pipelines are directly evaluated. The 16,000 fps baseline is excluded from this comparative analysis to prevent self-referential bias toward the reference signal. Across the 16 feature-frame-rate combinations evaluated within the 2000-8000 fps range used for performance recommendations, Kalman–sinc achieved lower error in 12 cases (75%) (Tables 2 – 3).

The advantage of Kalman smoothing was most evident for peak velocity features (POV and PCV), with the improvement increasing at higher frame rates. For Peak Relative Opening Velocity (POV), the reduction in MAPE achieved by Kalman–sinc compared with FFT–sinc increased from 4.2–4.9 percentage points (11–15% relative reduction) at 2000–4000 fps to 10.7–11.9 percentage points (35–44% relative reduction) at 5333–8000 fps. Peak Relative Closing Velocity (PCV) showed a similar trend, although with a smaller magnitude of improvement: Kalman–sinc reduced MAPE by 1.3–3.5 percentage points (4–11% relative reduction) at 2000–4000 fps and by 4.8–5.0 percentage points (17–18% relative reduction) at 5333–8000 fps.

The improvement was not universal across all features. Average Relative Closing Velocity (ACV) was an exception, with FFT–sinc consistently producing lower errors than Kalman–sinc over the evaluated frame-rate range. For example, at 8000 fps, ACV MAPE was 11.3% with FFT–sinc compared with 15.6% with Kalman–sinc. This suggests that the relative performance of the denoising methods depends on the specific characteristics of the kinematic feature being estimated, with Kalman smoothing providing the greatest benefit for features that are highly sensitive to localized noise amplification, such as peak velocity measurements.

## 4. **Discussion**

This study investigated the relationship between temporal sampling requirements and spatial resolution in high-speed videoendoscopy (HSV) to establish a frame-rate–resolution framework for accurate vocal fold kinematic measurement. The central finding is that temporal resolution cannot be optimized independently of spatial resolution: increasing frame rate improves sampling density only until the inter-frame displacement approaches the spatial quantization limit. Beyond this point, additional frames primarily capture quantization noise rather than additional biological motion. This phenomenon was formally derived in Section 2.1 via the bound $N \le K \min(S_q, S_q^{-1})$ and experimentally confirmed in section 3.2. Specifically, in our experimental dataset, spatial resolution was held constant while frame rate was scaled up to 16,000 fps; under these conditions, the upper theoretical sampling bound for the frame ($f_s$) was violated. Substituting for typical values for $K, S_q$ from our dataset yields theoretical upper bounds ($f_{s,max}$) as (~9400 fps, ~5100 fps) for females and males respectively. As a direct result, Mean Absolute Percentage Error (MAPE) for derivative-based features in the unprocessed waveforms degraded substantially with increasing frame rate. The following sections further discuss population-specific operating windows (Section 4.1), reconcile these experimental findings with theoretical predictions (Section 4.2), and evaluate optimal signal-processing strategies (Section 4.3).

### 4.1. Defining the "Operating Window" across the populations

The frame rate bounds derived in section 2.1 ($f_{s,\min}$) and ($f_{s,max}$) demonstrate that no single recording frame rate is optimal for all patient populations. Instead, the appropriate operating range depends on the interaction between the subject's vocal characteristics (fundamental frequency, open quotient, and speed quotient) and the available spatial resolution (pixel excursion, ($K$)).

Deliyski et al.[24] previously reported that a recording frame rate of 2000 fps is often inadequate for evaluating high-pitched voices. The analytical framework developed in this study quantifies this limitation and defines the corresponding operating window. The equations for the frame rate bounds ($f_{s,min}$, $f_{s,max}$) reveals that for unprocessed displacement waveform, the number of frames per phase ($N$) increases linearly with frame rate, while the inter-frame displacement decreases monotonically with frame rate. Figure 8 illustrates this trade-off by plotting both parameters across a range of frame rates for a representative child's voice. Valid kinematic reconstruction for unprocessed displacement waveform is constrained to the narrow operating region bounded by the vertical dashed lines [$f_{s,min}$, $f_{s,max}$] (6667 – 7111 Hz) where $N >= 5$ and inter-frame displacement remains above 1 pixel/frame. Figure 9 extends this analysis across demographics, illustrating how these operating windows shift under representative glottal parameters. For a representative female speaker ($f_0 = 220\ Hz, K = 14,\ S_q = 1.2, O_q = 0.6$ ), the predicted frame-rate bounds are approximately [4033, 9411] fps. Similarly, the range for a representative male ($f_0 = 120\ Hz, K = 14,\ S_q = 1.2, O_q = 0.6$ ), the frame rate bounds are [2200, 5133.3] fps. These operating windows nearly encompass the commonly used clinical recording rate of 4000 fps but excludes 16,000 fps. This prediction is consistent with the experimental results, which showed substantial degradation of derivative-based features at 16,000 fps in the absence of appropriate noise suppression (Table 3).

The constraints are even more restrictive for pediatric and other high-pitched voices. For a representative child ($f_0 = 320\ Hz, K = 8,\ S_q = 1.5, O_q = 0.6$ ), the predicted operating window narrows to an interval approximately of [6667 , 7111] fps (Figure 9). Under these conditions, the conventional 4000 fps recording rate provides insufficient temporal sampling, whereas increasing the frame rate toward 8000 fps without a corresponding increase in spatial resolution causes the inter-frame displacement to approach the quantization limit. As a result, derivative-based features become increasingly contaminated by quantization noise. In such scenarios, maximizing spatial resolution—whether through higher sensor pixel density or a tighter optical crop focused on the vocal folds—is essential to preserve measurement fidelity and widen the valid operating window.

These examples illustrate that the operating frame-rate window is population dependent and cannot be defined independently of image resolution. For high-pitched voices, increasing frame rate alone is not sufficient; adequate pixel excursion must first be achieved through improved spatial resolution or magnification. Otherwise, the potential gains in temporal sampling are offset by increased quantization error, particularly for velocity-based kinematic measurements.

### 4.2 The Spatial–Temporal Sampling Trade-off: Reconciling Temporal and Spatial Resolution

Consistent with the analytical framework's prediction that temporal differentiation amplifies spatial quantization noise, the experimental results demonstrated that the impact of the spatial–temporal sampling trade-off varied substantially across kinematic features based on their mathematical formulation.

Peak Lateral Displacement (PLD), an amplitude-based feature that does not depend on temporal differentiation, remained robust across all evaluated conditions. Similarly, the timing-based Open Quotient (OQ) was stabilized primarily through interpolation. In contrast, Speed Quotient (SQ) and derivative-based features, particularly peak velocity measurements, exhibited substantial degradation when frame rates exceeded the usable temporal sampling range under no-processing conditions. At 16,000 fps, Peak Relative Opening Velocity (POV) and Peak Relative Closing Velocity (PCV) errors reached 145.6% and 242.8% MAPE, respectively (Table 3), directly supporting the predicted upper frame-rate constraint.

Once the frame rate exceeds ($f_{s,max}$), interpolation alone cannot restore signal fidelity because the underlying samples no longer contain sufficient spatial information. When the inter-frame displacement becomes smaller than one pixel, the waveform becomes dominated by staircase quantization artifacts rather than the continuous vocal fold motion. This limitation was evident even with sinc interpolation alone, which produced a PCV MAPE of 231% at 16,000 fps. These findings demonstrate that increasing frame rate beyond the spatial-resolution-dependent ceiling does not increase measurement information; instead, it amplifies the influence of spatial quantization noise.

The derived sampling and noise constraints are applicable not only to full HSV recordings but also to displacement waveforms obtained from laryngeal videokymography. In full HSV, spatial processing of the vocal fold contour prior to edge localization may improve displacement estimation by incorporating information from neighboring pixels. Under an idealized uniform quantization noise model $U[-0.5, 0.5]$, averaging $W$ independent neighboring measurements reduces noise variance by a factor of $W$, suggesting that spatial smoothing could in principle lower the effective quantization noise.

However, the practical benefit of spatial smoothing is limited by two competing effects. First, quantization errors among spatially adjacent pixels are not necessarily independent, reducing the expected noise reduction. Second, excessive smoothing may combine pixels representing different physical locations along the vocal fold edge, producing contour blurring and displacement bias rather than true noise suppression. The magnitude of this tradeoff depends on local vocal fold curvature and the spatial extent of the intensity transition at the edge, factors not evaluated in the current study. Therefore, the derived frame-rate bounds should be interpreted as a conservative baseline for HSV recordings. Future studies should quantify the effective reduction in quantization noise achievable through two-dimensional contour-based smoothing and determine how this modifies the spatial sampling limit.

The upper frame rate limit ($f_{s,\max}$), is fundamentally determined by the relationship between inter-frame pixel displacement, frame rate, and peak pixel excursion ($K$). As illustrated in Figure 6, without advanced denoising, the Peak Relative Closing Velocity (PCV) error increased from 48.9% MAPE at 2000 fps to 242.8% at 16000 fps. This behavior is consistent with the analytical prediction that, once $f_s$ exceeds $f_{s,max}$, temporal oversampling no longer improves measurement accuracy and instead exposes spatial quantization as the dominant source of error. Thus, the effective temporal resolution of a HSV system is ultimately constrained by the spatial excursion available in the image: a higher frame rate is only useful when supported by sufficient pixel displacement.

For derivative-based kinematic features, interpolation alone is therefore insufficient. Because differentiation amplifies high-frequency noise, reliable velocity estimation requires both adequate temporal sampling and appropriate noise suppression before feature extraction.

### 4.3 Superiority of the Kalman-Sinc Pipeline over Spectral Filtering

The comparative evaluation (Section 3.2.3, Figure 7) identified Kalman smoothing combined with sinc interpolation as the most robust processing framework among all evaluated approaches across a cohort of 12 subjects ($N = 12$, 6 males, 6 females). Although FFT–sinc achieved marginally lower error in a minority of condition combinations (specifically for ACV), Kalman–sinc demonstrated superior performance across the vast majority of feature–frame-rate conditions, particularly for peak velocity parameters (POV and PCV). Selecting Kalman–sinc as the unified processing framework avoids the practical complexity of deploying feature-dependent pipelines while optimizing performance for peak kinematic features. Although sinc interpolation effectively exploits the global periodic structure of sustained vocal fold vibration, interpolation alone cannot recover information lost to spatial quantization. Interpolation reconstructs missing temporal samples accurately only when the underlying discrete measurements are adequately represented by available image data.

The advantage of Kalman smoothing arises from its incorporation of a dynamic model of the underlying system. Unlike conventional FFT-based denoising, which primarily suppresses frequency components without explicit consideration of vocal fold biomechanics, the Kalman filter operates within a state-space framework that models the displacement trajectory and its temporal evolution. By incorporating the expected continuity of vocal fold motion, the Kalman filter reduces pixel-level quantization fluctuations while preserving the underlying vibratory dynamics.

This combination of physically informed smoothing and high-fidelity temporal reconstruction was particularly beneficial for derivative-based features, which are highly sensitive to noise amplification. For example, at 8000 fps, the Kalman–sinc pipeline reduced Peak Relative Opening Velocity (POV) MAPE to 15.4%, compared with 45.3% using sinc interpolation alone. These findings indicate that interpolation and noise reduction address distinct limitations: interpolation restores temporal resolution, whereas Kalman smoothing suppresses spatial quantization artifacts that would otherwise be amplified during kinematic feature extraction.

Beyond temporal and spectral filtering, it is worth noting whether sub-pixel edge detection could obviate the need for temporal smoothing or relax the derived frame-rate limits. While sub-pixel algorithms reduce position step sizes, they evaluate frames independently and remain constrained by the low signal-to-noise ratio inherent to high-speed videoendoscopy. The short exposure times required at high frame rates necessitate spatial image smoothing, which blurs edge boundaries and shifts estimated positions, particularly near glottal closure and maximum lateral excursion, where frame-to-frame displacement drops below the sub-pixel precision threshold. In contrast, temporal filtering via Kalman smoothing incorporates vocal fold kinematics (position, velocity, and acceleration over time) across successive frames. Sub-pixel spatial localization and temporal Kalman smoothing are thus complementary: sub-pixel estimation improves spatial precision per frame, while temporal smoothing filters the trajectory across frames.

An alternative strategy to mitigate spatial quantization noise at high sampling rates is ensemble cycle averaging. While averaging across consecutive vibratory cycles reduces uncorrelated noise, it collapses the signal into a single mean cycle, eliminating cycle-to-cycle perturbations in amplitude, period, and velocity. Preserving these dynamic variations is a primary advantage of high-speed videoendoscopy over videostroboscopy, particularly in dysphonic or irregular phonation where automated cycle demarcation is unreliable. In contrast, the Kalman–sinc pipeline operates sequentially on a frame-by-frame basis, suppressing spatial quantization artifacts while retaining cycle-to-cycle kinematic information.

### 4.4 Evidence-Based Recommendations for Kinematic Feature Extraction

Based on the observed stabilization of feature errors across frame rates and processing conditions, we propose a feature-dependent strategy for HSV-based kinematic analysis. The results demonstrate that different kinematic features have distinct sensitivity to temporal sampling, spatial quantization, and noise-processing requirements.

Peak Lateral Displacement (PLD) was the most robust feature evaluated, maintaining relatively low error across all tested frame rates with minimal processing requirements. Timing-based features showed intermediate sensitivity. Open Quotient (OQ) and Speed Quotient (SQ) benefited from temporal reconstruction, with sinc interpolation providing improved stability at clinically relevant frame rates. To reduce potential sinc-ringing artifacts (Gibbs phenomenon)[34] associated with abrupt transitions during glottal opening and closing, a hybrid interpolation approach incorporating linear extrapolation at the phase boundaries was used. This approach allowed OQ estimation error to stabilize at approximately 4%.

Velocity-derived features represented the highest-sensitivity category and should not be extracted from minimally processed waveforms. Because differentiation amplifies high-frequency noise, reliable estimation of velocity features requires both noise suppression and temporal reconstruction. Based on the evaluated conditions, the Kalman smoothing followed by sinc interpolation pipeline at frame rates of approximately 8000 fps or higher provided the most consistent performance for velocity-based measurements. These features may therefore be more suitable for studies evaluating vocal fold biomechanics, stiffness, and neuromuscular control, where accurate characterization of motion dynamics is required.

The Kalman–sinc pipeline was selected as the recommended general processing framework based on its overall performance across the evaluated feature set (Tables 1–3). Although FFT–sinc occasionally produced slightly lower error for individual feature–frame-rate combinations, the differences were not considered sufficient to justify feature-specific processing recommendations given the limited sample size of the high-frame-rate dataset ($\boldsymbol{N} = \mathbf{12}$). Instead, Kalman–sinc is recommended as a single, broadly applicable processing standard that provides consistent performance across diverse kinematic measurements.

The recommended minimum frame rates and corresponding MAPE values obtained in this study are summarized in Table 4. As described in the methods section, all processing conditions listed in Table 4 (except for the "No Processing" condition) include upsampling to the 16000 fps reference grid before feature extraction. Therefore, the reported frame rates represent the native acquisition rates rather than the final temporal resolution used during analysis.

### 4.5. Implications for Clinical/Research Reporting and Future Standards

The findings of this study highlight the need for more comprehensive characterization and reporting of high-speed videoendoscopy (HSV) data. Reporting frame rate alone is insufficient for assessing the reliability of extracted kinematic measurements. We recommend that future studies and clinical/research reports include the peak pixel excursion ( $K$) associated with vocal fold lateral displacement alongside the acquisition frame rate and reported kinematic parameters.

As demonstrated analytically, the effective spatial resolution of the measurement system, quantified by ($K$), directly determines the feasible operating range of frame rates by defining the lower and upper sampling bounds ($f_{s,\min}$) and ($f_{s,\max}$). Without information regarding the number of pixels spanning the vocal fold displacement, the uncertainty and reliability of derived measures—particularly velocity and asymmetry-related features—cannot be fully interpreted. This observation is consistent with the empirical findings reported by Schlegel et al.[26] which emphasize the importance of spatial resolution in quantitative HSV analysis.

The observed improvement achieved with Kalman smoothing and sinc interpolation also supports the broader principle that kinematic feature stability depends on incorporating sufficient information from the underlying vibratory signal. By modeling the vocal fold trajectory across multiple cycles, the Kalman–sinc framework reduces sensitivity to individual-cycle noise and quantization fluctuations. This is consistent with previous work demonstrating improved parameter stability with longer analyzed sequences.[27] The datasets evaluated in the present study contained more than 20 vibratory cycles, satisfying this requirement. Furthermore, unlike cycle-by-cycle processing approaches, the Kalman–sinc pipeline incorporates temporal information across the available waveform, enabling a more robust estimate of the underlying periodic and quasi-periodic vocal fold motion.

### 4.6. Limitations and Future Work

This study was limited to a relatively small cohort of healthy adults ($N = 12$) for high-frame-rate kinematic evaluation. However, the narrower operating windows predicted for pediatric and high-pitched voices suggest that these populations may have an even greater need for state-space-based signal processing and optimized spatial-temporal sampling. Future work should therefore evaluate the proposed framework across larger and more diverse populations, including children and patients with voice disorders.

An important next step is extending the pipeline to highly aperiodic phonation. The primary challenge in such signals is expected to be sinc interpolation, which relies on the assumption of a quasi-periodic waveform. Significant cycle-to-cycle variations in amplitude and frequency may violate this assumption and introduce reconstruction artifacts. In contrast, the Kalman framework is expected to be more adaptable because it does not require strict periodicity and can incorporate the continuous dynamic structure of vocal fold motion. Even in disordered phonation, sufficient temporal organization may remain to support state-space estimation, including constant-jerk motion modeling, reliable identification of opening and closing phases, and appropriate parameter calibration.

For highly irregular signals, however, the frame-rate bounds derived in Appendix A should be applied conservatively. Rather than using a single representative cycle across the recording, the operating window should be determined from the most demanding vibratory cycle—specifically, the cycle with the highest fundamental frequency (shortest period) and the greatest degree of phase asymmetry (extreme speed quotient). This approach ensures that the sampling requirements remain valid under worst-case vibratory conditions.

Extending this framework from displacement-based waveforms to the Glottal Area Waveform (GAW) requires additional theoretical development. Unlike vocal fold displacement measurements, where signal amplitude and quantization noise exist within the same one-dimensional spatial coordinate system, GAW estimation involves a two-dimensional area measurement. The glottal area signal is determined by the enclosed region, whereas quantization noise originates primarily from pixels along the glottal boundary. Each boundary pixel contributes uncertainty to the estimated area, causing the total noise magnitude to depend on the number and configuration of perimeter pixels.

Consequently, GAW quantization noise is shape dependent: a glottal configuration with a complex or elongated boundary and a relatively large perimeter-to-area ratio will exhibit lower signal-to-quantization-noise ratio than a simpler shape with the same area. Therefore, direct application of the displacement-based noise model and frame-rate bounds to GAW is not appropriate without modification. Developing equivalent spatial noise models and frame-rate constraints for GAW represents an important but non-trivial extension of the present framework.

## 5. Conclusion

This study characterized the effects of temporal and spatial sampling limitations on vocal fold kinematic measurements and evaluated the strategies for reducing these errors through Kalman smoothing and temporal interpolation. We derived the analytical frame rate bounds as a function of pixel excursion, speed quotient and open quotient. These bounds define a minimum frame rate required to obtain $N$ samples within each vibratory phase and a maximum frame rate beyond which kinematic accuracy deteriorates due to spatial quantization effects. An online calculator for estimating these frame-rate limits is available at: https://hari-krishnan.github.io/hsdi-frame-rate-calculator.

A central finding of this work is that increasing frame rate does not always improve measurement precision. As the maximum usable frame rate is directly determined by the available pixel excursion, temporal and spatial resolution must be optimized together rather than considered independently. Our derived operating condition ( $N \leq K \min\left(S_q,\ {S_q}^{-1}\right)$) demonstrates that a valid operating interval for the frame rate exists only when the available spatial resolution is sufficient to support the desired temporal resolution and degree of cycle asymmetry. This constraint is particularly important for pediatric and other high-pitched voices, where higher fundamental frequencies increase the minimum frame rate ($f_{s,min}$) and may exceed the commonly used clinical recommendation of 4000 fps resulting in temporal undersampling. Conversely, an upper bound ($f_{s,min}$) governs low-frequency (adult male) voices operating with small displacement amplitudes, such as during soft or pressed phonation. At elevated frame rates, inter-frame displacement drops near or below the camera's spatial pixel quantization threshold at higher frame rates. To facilitate practical application, our open-access online calculator enables clinicians and researchers to establish population-specific recording boundaries for optimized HSV analysis.

To address these sampling limitations, we identified Kalman smoothing combined with sinc interpolation as the most effective processing framework. This approach separates the underlying kinematic signal from spatial quantization noise by combining physically informed noise reduction with high-fidelity temporal reconstruction. The resulting improvement was most pronounced for derivative-based features, which are inherently sensitive to noise amplification.

Based on our findings we recommend a stratified, feature-driven approach to clinical HSV processing: For peak lateral displacement, 2000 fps recordings with minimal processing are sufficient because this amplitude-based feature is relatively insensitive to temporal sampling limitations. For Open Quotient, use of sinc interpolation is recommended to improve temporal reconstruction and enable reliable estimation at frame rates as low as 2000 fps, whereas Speed Quotient demands the Kalman–sinc pipeline to mitigate cycle-asymmetry noise at the same frame rate. Conversely, derivative-based velocity and stiffness-related measures, demand the full Kalman-sinc pipeline combined with frame rates of approximately 8000 fps or higher.

It is recommended that clinicians and researchers should operate at or near the derived minimum rate whenever possible and report peak pixel excursion ($K$) as a standard acquisition parameter to contextualize the reliability of extracted kinematic measures. These reporting and frame rate guidelines provide a structured framework applicable to any kinematic features derived from 1D displacement waveforms across high-speed videoendoscopy (HSV), digital kymography (DKG), and videokymography (VKG). However, because these recommendations are derived from 1D displacement dynamics, they cannot be generalized beyond displacement-based parameters (e.g., to 2D area metrics, mucosal wave propagation, or phase asymmetry) without separate, parameter-specific validation. Adoption of such a spatial-temporal sampling framework is essential for reliable quantitative assessment of vocal fold biomechanics.

**Acknowledgments**

This work was funded by National Institutes of Health (NIH) - National Institute on Deafness and Other Communication Disorders Grants R01DC017923 and R03DC11360. The authors gratefully acknowledge Dr. Dimitar Deliyski for providing 16000 fps High-Speed data.

## Appendices

### Appendix A: Noise Floor modeling and Analytical Derivation of Sampling Bounds in Laryngeal High-Speed Imaging

From the High-Speed Videoendoscopy frames, the lateral displacement is computed for each frame by tracking the vocal fold edge position relative to the medial line. This process yields a discrete-time displacement waveform $x[n]$ sampled at intervals determined by the acquisition frame rate: $t_n = n/f_s$, where $f_s$ is the frame rate, $n = 0, 1, \ldots$ is the temporal frame number and $t_n$(seconds) represents the $n^{th}$ sampling instant. During displacement tracking, two primary digital sampling limitations influence measurement accuracy. The first is spatial quantization, which arises because continuous vocal fold motion is represented on a discrete pixel grid. The second is temporal sampling, which determines the number and timing of observations available to capture the vocal fold edge motion within each vibratory cycle. Together, these effects establish a fundamental noise floor and limit the accuracy of derived kinematic features.

This appendix first provides the mathematical derivation of the minimum (min) frame rate ($f_{s,min}$) and maximum (max) frame rate ($f_{s,max}$) bounds, required for accurate and efficient sampling of vocal fold displacement waveforms. The derivation incorporates glottal cycle characteristics, including fundamental frequency, open quotient, speed quotient, and spatial excursion. The resulting analysis demonstrates the intrinsic coupling between temporal sampling rate and spatial resolution and establishes the conditions under which reliable kinematic measurements can be obtained.

### Derivation of frame-rate bounds

We derive the analytical frame-rate bounds that satisfy both temporal and spatial sampling constraints as a function of clinically relevant vocal fold vibratory cycle characteristics. A quasi-periodic vocal fold vibratory cycle can be described by three primary parameters (figure 1):

- Fundamental Frequency ($f_0$): The reciprocal of the total glottal cycle time period ($T_0$)
- Open Quotient ($O_q$): The ratio of the total open-phase duration ($T_{open}$) to the total period ($T_0$)
- Speed Quotient ($S_q$): The ratio of the opening phase duration ($T_{opening}$) to the closing phase duration ($T_{closing}$).

Throughout this derivation, $f_0$ and frame rate ($f_s$) are in Hertz (Hz) (cycles per second and frames per second, respectively), and all temporal durations $T$ are in seconds. Open and closing quotients are unit less ratios.

This gives us Open Quotient ($O_q$) as:

$$O_q = \frac{T_{opening} + T_{closing}}{T_0} = \left(T_{opening} + T_{closing}\right) f_0 \qquad (1)$$

Similarly Speed Quotient $S_q$ as:

$$S_q = \frac{T_{opening}}{T_{closing}} \tag{2}$$

From these baseline definitions, the absolute durations of the opening and closing phases can be explicitly isolated. Using the speed quotient relation:

$$T_{opening} = S_q \times T_{closing} \tag{3}$$

Substituting this term (3) back into the open quotient expression (1) and solving for $T_{closing}$:

$$T_{closing} = \frac{O_q}{(S_q + 1) \times f_0} \tag{4}$$

Subsequently, the opening phase duration can be derived by substituting for $T_{closing}$ from (3) to (1) as below:

$$T_{opening} = \frac{O_q S_q}{(S_q + 1) \times f_0} \tag{5}$$

**Derivation of the Analytical Minimum Bound for frame rate**

Let $N$ be the minimum samples per phase required for reliable estimation of the derivative based kinematic features. Because the opening and closing phase durations can be asymmetric ($S_q \neq 1$), the minimum temporal sampling rate must satisfy the criteria for the shorter of the two phases ($T_{short}$):

$$f_s \geq \frac{N}{T_{short}} \tag{6}$$

$$f_{s,min} = \frac{N}{T_{short}} \tag{7}$$

Where the duration of the shorter phase is defined as:

$$T_{short} = \min\left(T_{opening}, T_{closing}\right) \tag{8}$$

Where min is the minimum operator. Substituting the isolated phase expressions derived in equations 4, 5 and simplifying:

$$T_{short} = \min\left(\frac{O_q S_q}{(S_q + 1) \times f_0}, \frac{O_q}{(S_q + 1) \times f_0}\right) \tag{9}$$

Taking the common terms out of the min operation:

$$T_{short} = \frac{O_q}{(S_q + 1) \times f_0} \min(1, S_q) \tag{10}$$

By substituting for $T_{short}$ in equation 7, the analytical minimum frame rate bound is derived as:

$$f_{s,min} = \frac{N(S_q + 1)\, f_0}{O_q \min(S_q,\ 1)} \tag{11}$$

**Derivation of the Analytical Maximum Bound for frame rate**

The upper limit on the usable frame rate is determined by the spatial resolution of the imaging system. In digital edge tracking, the smallest measurable displacement is constrained by the pixel grid, where a quantization step size of $\Delta = 1$ pixel introduces an inherent spatial measurement uncertainty. Let $K$ denote the peak glottal excursion expressed in discrete pixel units.

If the frame rate is increased without a corresponding increase in spatial magnification, the physical displacement traversed by the vocal fold between consecutive frames drops below 1 pixel. Once the inter-frame displacement becomes smaller than one pixel, successive measurements no longer provide independent spatial information and the displacement waveform becomes dominated by staircase quantization effects. Under these conditions, numerical differentiation required for estimating velocity and acceleration parameters becomes increasingly sensitive to quantization noise and cannot reliably represent the underlying continuous motion without additional processing.

To prevent excessive temporal oversampling, the mean pixel displacement between consecutive frames ($\Delta x$) during the longest (slowest) vibratory phase ($T_{long}$) is constrained to remain at least 1 pixel.

$$T_{long} = \max(T_{opening}, T_{closing}) = \frac{O_q}{(S_q + 1) \times f_0} \max(1, S_q) \tag{12}$$

On an average during the longest phase the vocal fold edge moves $v$ pixels per second where:

$$v = \frac{K}{T_{long}} \tag{13}$$

The discrete inter-frame displacement as given by displacement velocity relation is:

$$\Delta x = \frac{v}{f_s} = \frac{K}{f_s\, T_{long}} \tag{14}$$

Where $K$ is the zero-to-peak amplitude in pixels. This is a linear approximation of the displacement during the longest phase. Setting the boundary constraint $\Delta x \geq 1$pixel yields:

$$\Delta x = \frac{K}{f_s T_{long}} \geq 1 \tag{15}$$

Implying:

$$f_s \leq \frac{K}{T_{long}} \tag{16}$$

$$f_{s,max} = \frac{K}{T_{long}} \tag{17}$$

Rearranging terms after substituting for $T_{long}$ from equation 12 and isolating the final expression for the analytical maximum frame rate bound:

$$f_{s,max} = \frac{K f_o (1 + S_q)}{O_q \max(S_q, 1)} \tag{18}$$

**Conclusion**

In this appendix we have demonstrated a complete algebraic derivation for the optimal frame-rate bound given by equations 11 and 18. Combining the inequalities we get:

$$\frac{N(S_q + 1)\, f_0}{O_q \min(S_q,\ 1)} \leq f_s \leq \frac{K f_o (1 + S_q)}{O_q \max(S_q, 1)} \tag{19}$$

Implying:

$$\frac{N(S_q + 1)\, f_0}{O_q \min(S_q,\ 1)} \leq \frac{K f_o (1 + S_q)}{O_q \max(S_q, 1)} \tag{20}$$

Canceling the common terms:

$$N \leq \frac{K \ \min(S_q, 1)}{\max(S_q,\ 1)} \tag{21}$$

For $S_q > 1$, equation (21) becomes $N \leq \frac{K}{S_q}$ and for $S_q < 1$, equation (21) becomes $N \leq KS_q$. Condensing both into expression gives:

$$N \leq K \min(S_q, 1/S_q) \tag{22}$$

Implying that for higher sampling rates as $N$ increases, a corresponding increase in spatial resolution (to increase $K$) is required to keep the digital sampling of the vocal fold motion ideal.

In summary, $N$ the minimum number of samples required per phase is determined by the requirement to compute the derivative accurately. $N$ and $T_{short}$ together decide the minimum frame rate needed $f_{s,min}$. $K$ the maximum pixel excursion or the effective resolution, determines the upper bound $f_{s,max}$. Enforcing $f_{s,min} < f_{s,max}$ gives a range of feasible frame rates and enforces a relationship between $N$ and $K$, for the given subject specific parameters.

## Appendix B: Kalman Smoother for Lateral Displacement from Laryngeal High-Speed Videoendoscopy

The Kalman smoother is a recursive Bayesian estimator that produces optimal state estimates by combining a dynamic motion model with noisy observations from the full measurement sequence[29]. It is applied here to the raw lateral displacement waveform extracted from High-Speed Videoendoscopy (HSV) frames, to suppress the sampling noise while preserving the rapid kinematic transitions.

### Formulation

The state of vocal fold motion, the state vector, at temporal index $n$ is defined as: $X_n = [d_n, v_n, a_n, j_n]^T$, comprising lateral displacement from glottal midline ($d_n$), relative velocity ($v_n$), relative acceleration ($a_n$), and rate of change of relative acceleration (relative jerk) ($j_n$). The inclusion of jerk is critical as it allows the filter to capture the rapid force changes inherent in vocal fold collision while distinguishing genuine motion from stochastic jumps caused by pixel-level tracking errors. Although jerk continuously varies across a full vibratory cycle, treating $j_n$ as piecewise constant over individual sampling intervals introduces negligible approximation error compared to spatial pixel noise. This local frame-to-frame formulation accommodates time-varying cycle dynamics such as sinusoidal or asymmetric motion, without imposing rigid global harmonic assumptions.

The predicted state ($\hat{X}_n$) is propagated forward by one sampling interval ($T_s$) in seconds using the constant-jerk kinematic transition matrix ($F$):

$$\hat{X}_n = FX_{n-1}, \qquad F = \begin{bmatrix} 1 & T_s & \frac{T_s^2}{2} & \frac{T_s^3}{6} \\ 0 & 1 & T_s & \frac{T_s^2}{2} \\ 0 & 0 & 1 & T_s \\ 0 & 0 & 0 & 1 \end{bmatrix}$$

where $X_{n-1}$, is the state at the previous time index ($n-1$). Each row of $F$ encodes how displacement, velocity, acceleration, and jerk at frame $n-1$ contribute to the corresponding state at frame $n$. The predicted state error covariance ($P_n$) is propagated as:

$$P_n = FP_{n-1}F^T + Q$$

The process noise covariance $Q$ is set to zero, reflecting the assumption that the constant-jerk kinematic model exactly describes inter-frame motion and that all estimation uncertainty arises from observation noise $r$. The Kalman gain ($K_n$) is then computed as:

$$K_n = P_n H^T (H\, P_n H^T + r)^{-1}$$

The state is subsequently corrected by the observation residual:

$$X_n = \hat{X}_n + K_n(Z_n - H\hat{X}_n)$$

The observation model ($Z_n$) is:

$$Z_n = H\mathrm{X}_\mathrm{n} + w_n, H = [1 \quad 0 \quad 0 \quad 0]$$

where observation matrix ($H$) indicates that only displacement is directly observable and $w_n$ is the observation noise. If the sole noise source is spatial quantization, its variance ($\tilde{r}$) is theoretically equal to $r = 1/12$, derived from the uniform quantization noise floor of a one-pixel step.

The filter is run in both the forward and backward temporal directions. The smoothed estimate is taken as the average of the two passe. Bidirectional smoothing reduces the elevated estimation uncertainty near the glottal opening and closing events, which arises from the absence of observations during the closed phase.

**Closed-phase handling**

When the vocal folds are completely closed, no glottal edge is detected, and displacement is set to zero by the tracking algorithm. To prevent these spurious observations from interfering with the estimates, during closed-phase samples the Kalman gain $\boldsymbol{K_n}$ is forced to zero. This causes the filter to propagate the state purely by prediction and maintain the momentum of the vocal fold trajectory. When the folds reopen, the next non-zero observation triggers the resumption of normal measurement updates from the current predicted state. This treatment prevents the ringing and overshoot artifacts that arise with spline interpolation or Butterworth filtering when they encounter the sharp amplitude discontinuity at glottal closure.

**Adaptive parameterization of observation noise variance**

The observation noise variance $\tilde{r}$ quantifies the reliability of each displacement observation relative to the kinematic model prediction. The theoretical value $r = 1/12$ corresponds to pure spatial quantization noise; the expected observation uncertainty when the edge is tracked precisely at the pixel grids without any sub-pixel estimation.

For a given recording, $\tilde{r}$ is selected via a geometric search initialised at $\tilde{r} = 1$ and decreased by a factor of 0.1 per step. At each step, the absolute error $e = |d_n - z_n|$ is computed for all open-phase samples where $z_n \neq 0$, and the search terminates when the mean absolute error $\bar{e}$ falls below 0.5 pixels, corresponding to the half-width of the quantization error interval $[-0.5, 0.5]$. Any further decrease in $\tilde{r}$ could artificially smooth the waveform.

Across all recordings, the search procedure converged to consistent $\tilde{r}$ values at each frame rate ($\tilde{r} = 10^{-5}$ at 2000 and 4000 fps; $\tilde{r} = 10^{-4}$ at 5333 fps; $\tilde{r} = 10^{-3}$ at 8000 fps and $\tilde{r} = 0.1$ at 16000 fps). At lower frame rates estimated $\tilde{r}$ values consistently fell below the theoretical value indicating that the edge detection algorithm's utilization of 2D spatial intensity gradients to localize vocal fold edges at a sub-pixel scale. The consistent increase in the observation noise $\tilde{r}$ with increase in frame rates demonstrates the high-speed CMOS sensor trade-off between temporal resolution and the uncertainty in the pixel intensity at ultra-high frame rates. It can be hypothesized that for VKG where the sub-pixel edge detection is restricted to 1D pixel line, observation noise variance $\tilde{r}$ will be closer to the $r$ (1/12).

**Sub-Sample Closed-Phase Boundary Detection**

Because the edge-tracking algorithm cannot follow the vocal folds during closure, the glottal opening and closing instants cannot be read directly from the displacement waveform. The Kalman-smoothed displacement $d_n = d_L[n] - d_R[n]$, defined as the difference between left and right vocal fold positions at frame $n$, is positive during the open phase and crosses zero at the opening and closing instants. Since estimation uncertainty is elevated near these transitions, sub-sample resolution is obtained by linear extrapolation: the sample straddling the zero-crossing is discarded, and the two immediately adjacent open-phase samples—preceding the closing instant or succeeding the opening instant— are used to extrapolate to $d = 0$. This yields sub-frame estimates of the opening and closing instants required for accurate computation of the open quotient and speed quotient.

## Tables

Table 1. Pooled mean absolute percentage error (MAPE, %) of Peak Lateral Displacement, Open Quotient, and Speed Quotient, with Kalman-smoothed 16000 fps recordings as reference, stratified by frame rate and processing pipeline. Lower values indicate closer agreement with the reference. (N = 12; 6 males, 6 females)

| Processing | Peak Lateral Displacement (PLD) | | | | | Open Quotient (OQ) | | | | | Speed Quotient (SQ) | | | | |
|---|---|---|---|---|---|---|---|---|---|---|---|---|---|---|---|
| | **2000 fps** | **4000 fps** | **5333 fps** | **8000 fps** | **16000 fps** | **2000 fps** | **4000 fps** | **5333 fps** | **8000 fps** | **16000 fps** | **2000 fps** | **4000 fps** | **5333 fps** | **8000 fps** | **16000 fps** |
| None | 9.35 | 7.33 | 9.86 | 8.57 | 4.23 | 12.96 | 3.52 | 14.95 | 8.55 | 7.28 | 70.39 | 33.34 | 23.05 | 11.46 | 19.94 |
| Spline only | 9.04 | 7.25 | 9.29 | 8.47 | 4.23 | 5.37 | 3.74 | 4.16 | 4.34 | 7.28 | 13.69 | 10.51 | 9.28 | 9.66 | 19.94 |
| Sinc only | 9.41 | 7.1 | 9.58 | 8.79 | 4.02 | 4.93 | 3.68 | 3.8 | 6.41 | 7.26 | 12.98 | 10.57 | 10.44 | 10.04 | 12.79 |
| FFT + Spline | 10.07 | 10.64 | 10.43 | 7.16 | 4.06 | 9.67 | 9.51 | 6.3 | 6.03 | 4.15 | 17.96 | 15.93 | 14.62 | 11.24 | 8.17 |
| FFT + Sinc | 10.2 | 10.76 | 10.29 | 7.14 | 4.06 | 9.21 | 9.64 | 6.39 | 5.87 | 4.15 | 17.19 | 15.14 | 14.83 | 11.03 | 8.26 |
| Kalman + Spline | 13.64 | 10.38 | 14.68 | 7.92 | 0 | 6.65 | 4.89 | 4.08 | 4 | 0 | 17.23 | 11.86 | 11.37 | 10.04 | 0 |
| Kalman + Sinc | 13.53 | 10.25 | 14.68 | 7.74 | 0.01 | 6.2 | 5.46 | 4.08 | 3.86 | 0.05 | 16.57 | 9.98 | 11.37 | 8.66 | 0.9 |

Table 2. Pooled mean absolute percentage error (MAPE, %) of Average Relative Opening and Closing Velocities with Kalman-smoothed 16000 fps recordings as reference, stratified by frame rate and processing pipeline. Lower values indicate closer agreement with the reference. (N = 12; 6 males, 6 females)

| Processing | Average Relative Opening Velocity (AOV) | | | | | Average Relative Closing Velocity (ACV) | | | | |
|---|---|---|---|---|---|---|---|---|---|---|
| | 2000 fps | 4000 fps | 5333 fps | 8000 fps | 16000 fps | 2000 fps | 4000 fps | 5333 fps | 8000 fps | 16000 fps |
| None | 33.55 | 18.43 | 14.9 | 13.04 | 15.92 | 37.38 | 23 | 20.84 | 17.98 | 9.93 |
| Spline only | 14.88 | 15.37 | 20.49 | 25.79 | 15.92 | 11.46 | 16.55 | 17.04 | 15.64 | 9.93 |
| Sinc only | 16.46 | 21.47 | 18.32 | 22.91 | 11.93 | 11.26 | 16.37 | 19.35 | 16.58 | 12.93 |
| FFT + Spline | 19.73 | 21.27 | 17.16 | 13.75 | 10.38 | 14.71 | 15.04 | 14.22 | 10.88 | 9.16 |
| FFT + Sinc | 20.42 | 21.66 | 17.31 | 13.93 | 10.29 | 15.04 | 14.57 | 14.26 | 11.27 | 9.37 |
| Kalman + Spline | 16.77 | 13.62 | 16.91 | 8.29 | 0 | 16.77 | 19.05 | 19.28 | 15.57 | 0 |
| Kalman + Sinc | 17.57 | 11.97 | 16.91 | 8.91 | 0.44 | 16.41 | 19.02 | 19.28 | 15.62 | 0.6 |

Table 3. Pooled mean absolute percentage error (MAPE, %) of Peak Relative Opening and Closing Velocities with Kalman-smoothed 16000 fps recordings as reference, stratified by frame rate and processing pipeline. Lower values indicate closer agreement with the reference. (N = 12; 6 males, 6 females).

| Processing | Peak Relative Opening Velocity (POV) | | | | | Peak Relative Closing Velocity (PCV) | | | | |
|---|---|---|---|---|---|---|---|---|---|---|
| | 2000 fps | 4000 fps | 5333 fps | 8000 fps | 16000 fps | 2000 fps | 4000 fps | 5333 fps | 8000 fps | 16000 fps |
| None | 50.05 | 34.3 | 26.38 | 37.39 | 145.63 | 48.89 | 37.04 | 51.86 | 75.39 | 242.77 |
| Spline only | 29.54 | 27.41 | 76.35 | 145.31 | 145.63 | 28.03 | 26.84 | 87.47 | 188.62 | 242.77 |
| Sinc only | 29.34 | 19.17 | 34.77 | 45.29 | 106.6 | 26.59 | 25.25 | 40.84 | 78.44 | 231 |
| FFT + Spline | 35 | 33.69 | 30.38 | 27.9 | 33.33 | 29.13 | 30.59 | 28.3 | 27.08 | 32.98 |
| FFT + Sinc | 36.69 | 33.69 | 30.71 | 27.31 | 33.35 | 30.73 | 31.54 | 28.47 | 27.03 | 32.98 |
| Kalman + Spline | 34.74 | 29.17 | 20.04 | 21.44 | 0 | 31.07 | 28.3 | 23.65 | 25.92 | 0 |
| Kalman + Sinc | 32.5 | 28.78 | 20.04 | 15.38 | 1.2 | 29.48 | 28.04 | 23.65 | 22.08 | 1.52 |

Table 4. Evidence-based recommendations for High-Speed Videoendoscopy (HSV) kinematic feature extraction, including recommended processing pipelines and minimum frame rate thresholds for processed versus unprocessed displacement waveforms. The Mean Absolute Percentage Error (MAPE) observed at the recommended processing is included in parenthesis. Except for the "No Processing" condition, all processing pipelines include upsampling to the 16,000 fps reference grid prior to feature extraction.

| Feature | Recommended Processing | Min. frame rate (fps) processing | Min. frame rate (fps) no processing |
|---|---|---|---|
| Peak Lateral Displacement | No Processing | Not Recommended | 2000 (9.35%) |
| Open Quotient | Sinc Interpolation | 2000 (4.93%) | 4000 (3.52%) |
| Speed Quotient | Kalman + Sinc | 2000 (9.98%) | 8000 (11.46%) |
| Average Relative Opening Velocity | Kalman + Sinc | 8000 (8.91%) | Not Recommended |
| Average Relative Closing Velocity | Kalman + Sinc | 8000 (15.62%) | Not Recommended |
| Peak Relative Opening Velocity | Kalman + Sinc | 8000 (15.38%) | Not Recommended |
| Peak Relative Closing Velocity | Kalman + Sinc | 8000 (22.08%) | Not Recommended |

## Figures

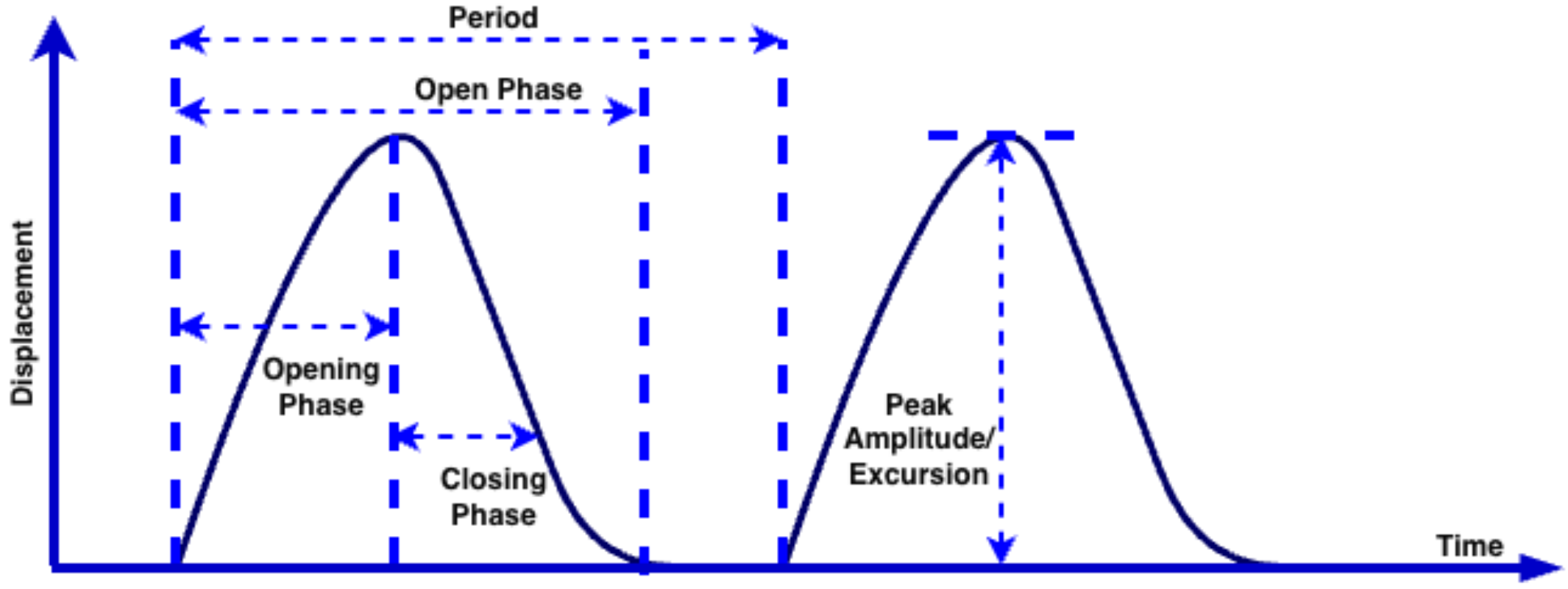

Figure 1: Illustration of a representative displacement waveform with key timing parameters marked.

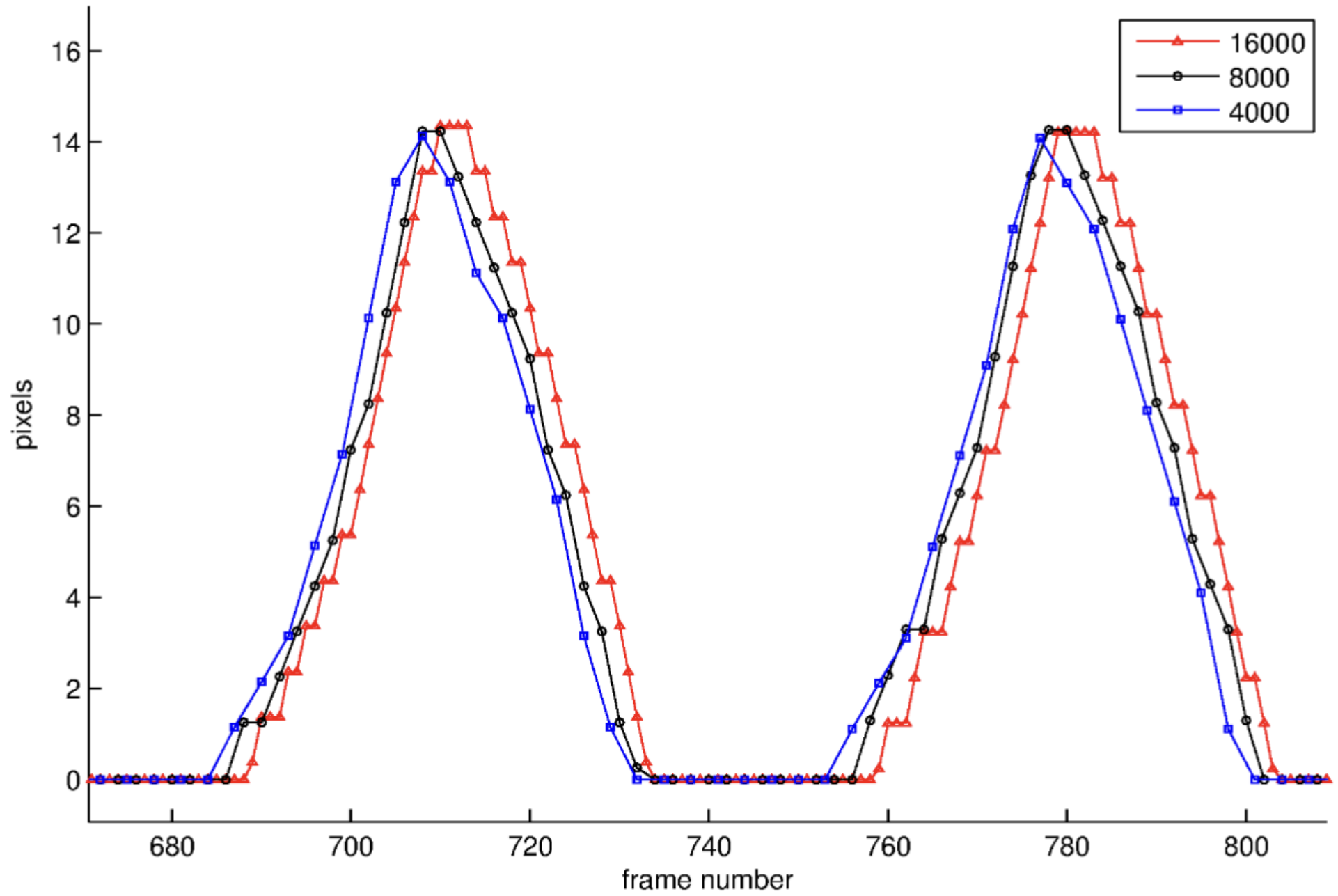

Figure 2. Displacement waveform (pixels) vs. frame number at three frame rates in frames per second (fps) (16000 fps, 8000 fps, 4000 fps). At 16000 fps the upper bound for the frame rate is often exceeded and inter frame displacement becomes < 1 pixel, so spatial quantization dominates and stair-stepping is evident; at 4000 fps temporal undersampling dominates (fewer samples per cycle).

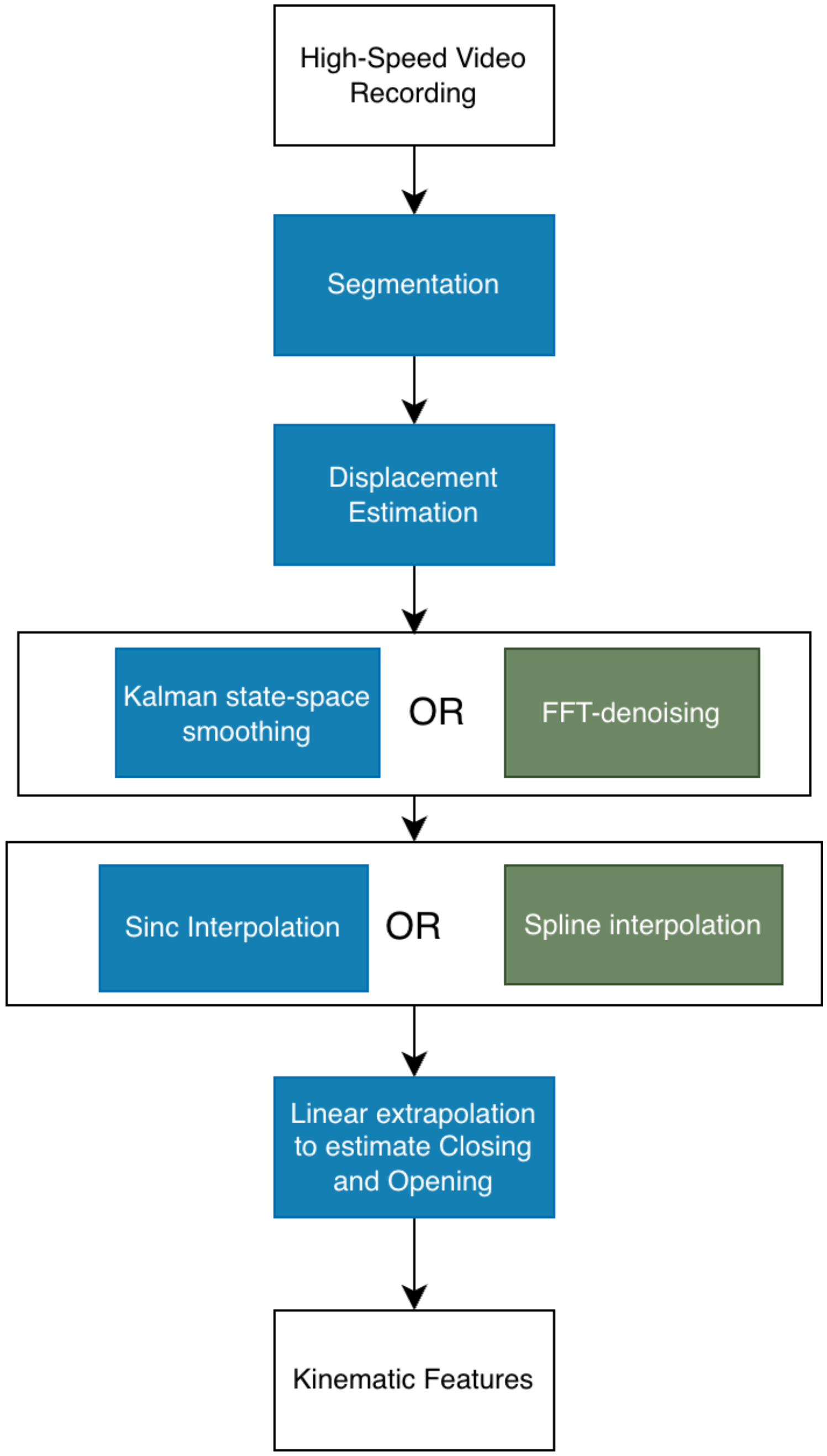

Figure 3. Post-processing pipeline for vocal fold displacement waveforms extracted from HSV recordings. Following glottal segmentation and kinematic extraction, waveforms undergo a smoothing stage — either Kalman smoothing or Fast Fourier Transform (FFT)-denoising — followed by temporal interpolation (spline or sinc) to enhance sampling density and velocity-estimate precision. The opening and closing phases are further refined via linear extrapolation to precisely localize the instants of glottal opening and closure.

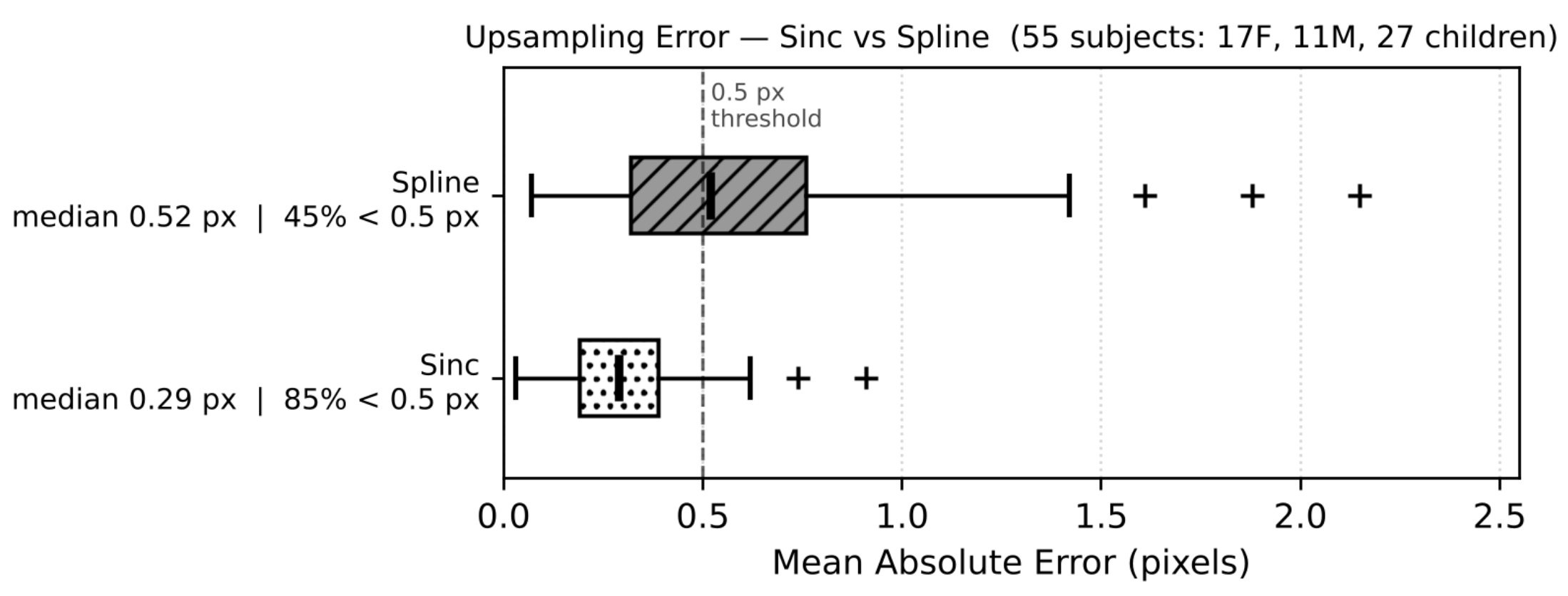

Figure 4. Comparison of reconstruction error between Sinc and Spline interpolation across downsampled frame rates (N = 55). High-speed video recordings originally acquired at 4,000 fps were downsampled to lower frame rates and subsequently reconstructed to evaluate temporal upsampling accuracy. Box plots display the mean absolute reconstruction error (in pixels) for Sinc and Spline interpolation methods across the study cohort (17 females, 11 males, 27 children). The vertical dotted line at 0.5 pixels indicates the practical accuracy threshold for glottal boundary detection; errors to the left of this line represent sub-pixel accuracy. Boxes indicate the interquartile range (IQR), internal bars denote medians, whiskers extend to 1.5 IQR and '+' symbols mark individual outliers.

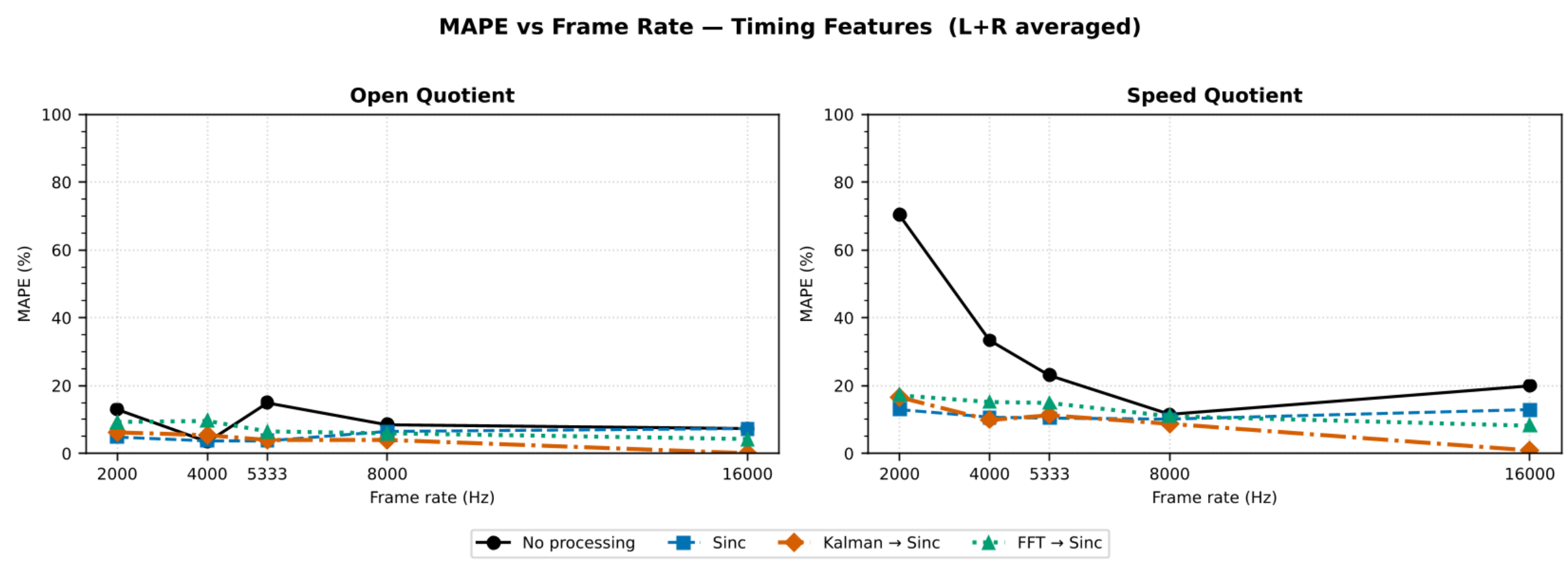

Figure 5. Mean Absolute Percentage Error (MAPE) of temporal glottal features (Open Quotient and Speed Quotient) as a function of frame rate (fps). Errors are calculated relative to the 16000 fps Kalman-smoothed reference (N=12 subjects: 6 females, 6 males). Feature values represent the average of the left and right vocal folds. The plot compares performance across four distinct processing pipelines: No processing (Raw), Sinc, Kalman + Sinc, and Fast Fourier Transform (FFT) + Sinc.

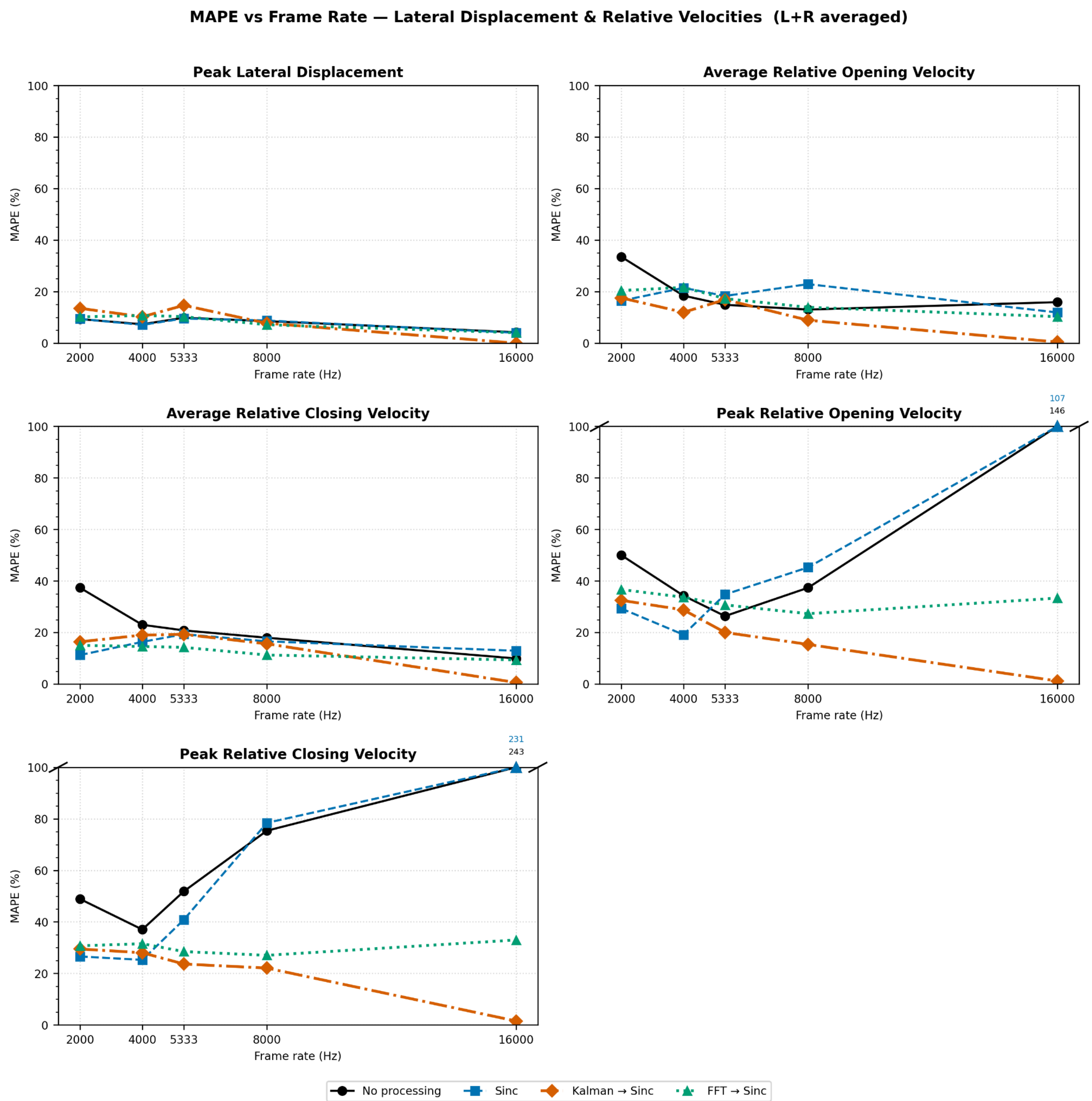

Figure 6. Mean Absolute Percentage Error (MAPE) of Peak Lateral Displacement and relative velocity features as a function of frame rate (fps). Displacement and velocity estimates are pooled (averages of left and right vocal folds) across four processing pipelines. To facilitate direct comparison, all panels utilize a uniform y-axis scale (0–100%); data points exceeding 100% are denoted by upward-pointing markers ( ▲ ) and labeled with their respective values. Errors are calculated relative to the 16000 fps Kalman-smoothed reference (N=12 subjects: 6 males, 6 females). The plot compares performance across four distinct processing pipelines: No processing (Raw), Sinc, Kalman + Sinc, and Fast Fourier Transform (FFT) + Sinc.

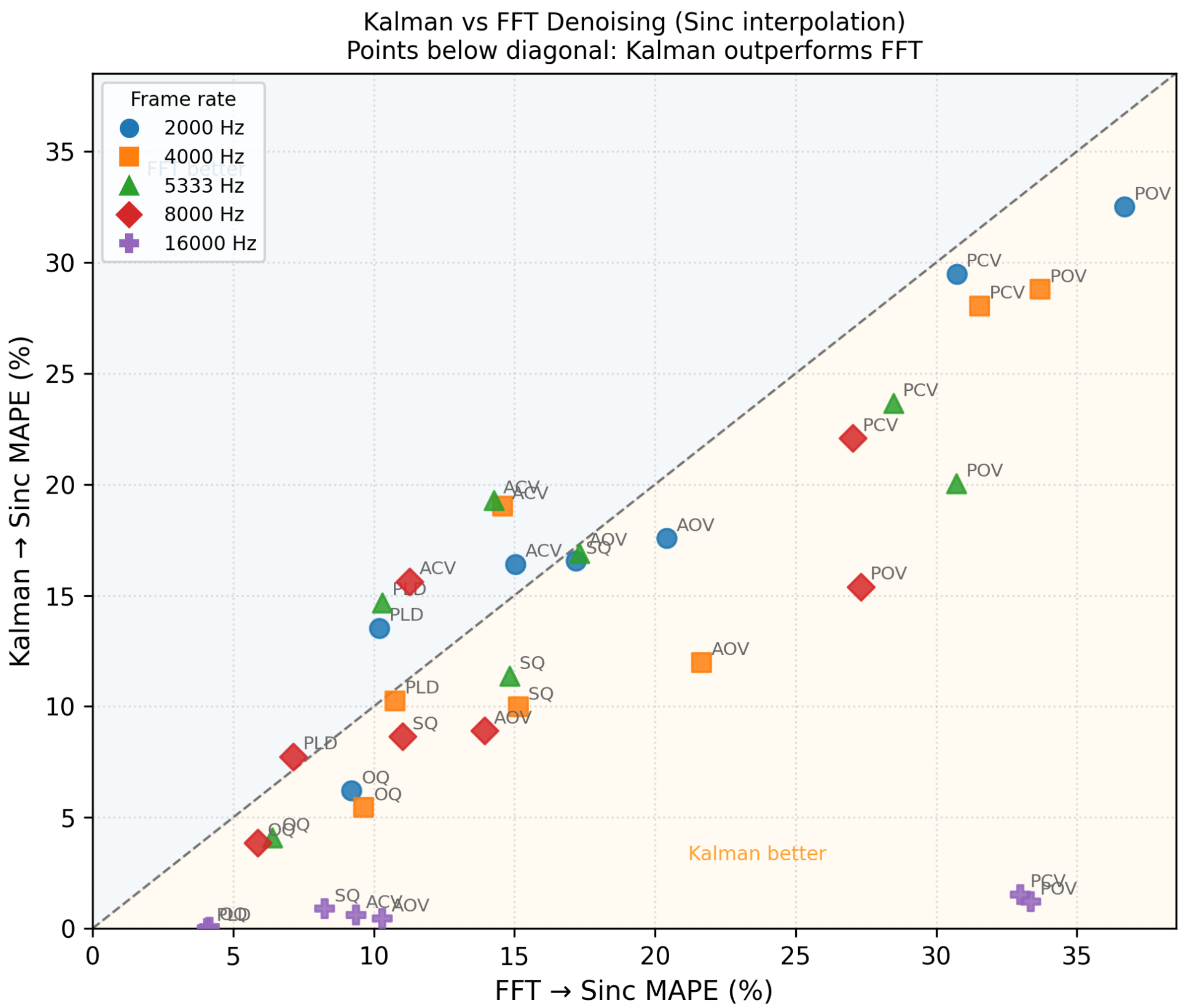

Figure 7. Scatter plot comparing Mean Absolute Percentage Error (MAPE) for the Kalman + Sinc and Fast Fourier Transform (FFT) + Sinc pipelines across seven kinematic feature classes at four frame rates (2000–8000 fps). Each data point represents a specific feature class at a given frame rate, with marker shape and color encoding temporal resolution. Points positioned below the unity diagonal indicate that Kalman smoothing outperforms FFT-denoising. All MAPE values are calculated relative to the 16000 fps Kalman-smoothed reference (N=12 subjects: 6 males, 6 females). The plot evaluates the following features: Peak Lateral Displacement (PLD), Open Quotient (OQ), Speed Quotient (SQ), Average Relative Opening Velocity (AOV), Average Relative Closing Velocity (ACV), Peak Relative Opening Velocity (POV), and Peak Relative Closing Velocity (PCV).

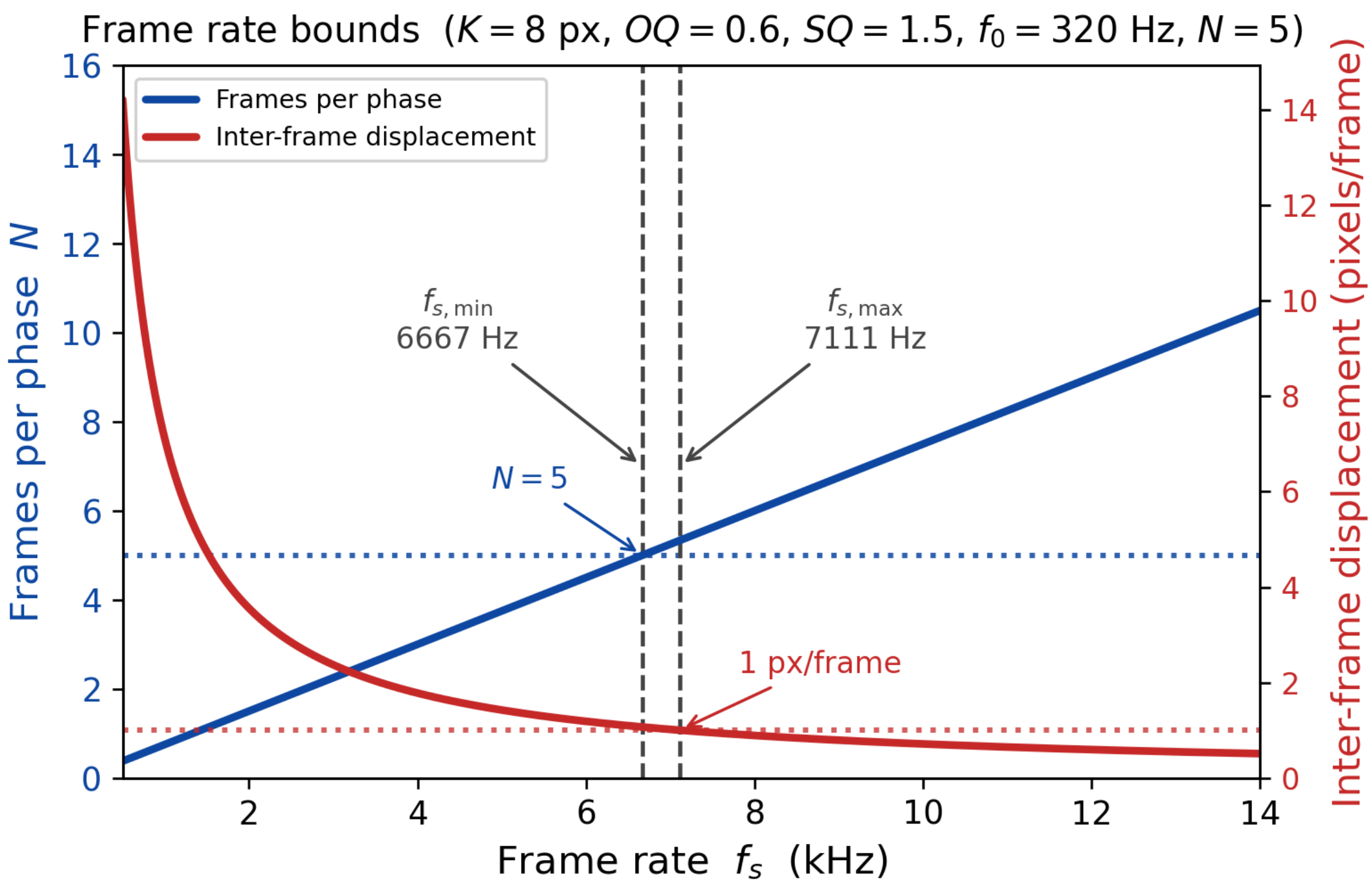

Figure 8. Variation of kinematic parameters with operating frame rate ($fs$) for unprocessed displacement waveform. Representative parameters for a child are used: pixel excursion $K$ = 8 pixels, open quotient ($OQ$) = 0.6, speed quotient ($SQ$) = 1.5, fundamental frequency ($f_0$) = 320 Hz, and minimum required frames per phase $N$ = 5. The number of frames per phase ($N$; blue line, left axis) increases linearly with frame rate, while the inter-frame displacement (red line, right axis) decreases monotonically. Valid kinematic feature estimation for unprocessed displacement waveform is constrained to the narrow operating region bounded by the vertical dashed lines [$f_{s,min}$, $f_{s,max}$] (6667 – 7111 Hz), where $f_{s,min}$ is the minimum frame rate needed to maintain $N$ >= 5 and $f_{s,max}$ is the maximum frame rate before inter-frame displacement falls below 1 pixel/frame.

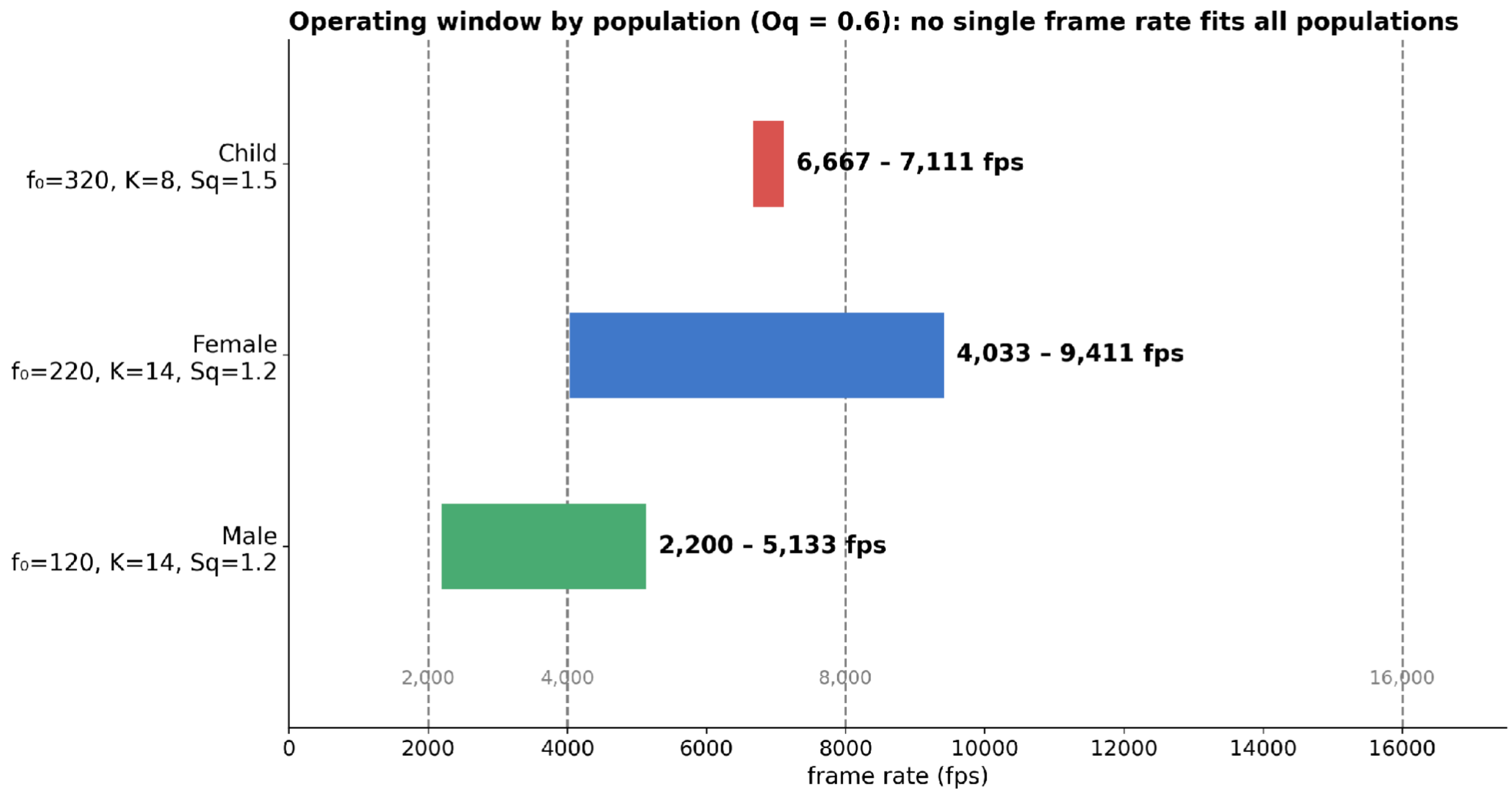

Figure 9. Operating window by population: no single frame rate fits all populations under representative glottal displacement waveform parameters: open quotient ($O_q$) = 0.6. The valid operating frame rate boundaries [$f_{s,min}$, $f_{s,max}$], where $f_{s,min}$ is the minimum operating frame rate boundary and $f_{s,max}$ is the maximum operating frame rate boundary, differ for each demographic: Child (fundamental frequency $f_0$ = 320 Hz, pixel excursion $K$ = 8 pixels, speed quotient $S_q$ = 1.5) is constrained to 6667 – 7111 fps, Female ($f_0$ = 220 Hz, $K$ = 14 pixels, $S_q$ = 1.2) spans 4,033 – 9,411 fps, and Male ($f_0$ = 120 Hz, $K$ = 14 pixels, $S_q$ = 1.2) ranges from 2200 to 5133 fps. Children have considerably narrower operating window compared to the adults.